\documentclass[10pt, conference, twocolumn, letterpaper]{IEEEtran}
\usepackage{booktabs}
\usepackage{amsmath}
\usepackage{amssymb}
\usepackage{graphicx}
\usepackage{fixmath}
\usepackage{algorithm}
\usepackage{algpseudocode}  
\usepackage{dsfont}
\usepackage{nth}
\usepackage{color,soul}
\usepackage{hyperref}
\usepackage{multirow}
\usepackage{comment}
\usepackage{balance}
\usepackage{paralist}
\usepackage{dblfloatfix}
\usepackage{siunitx}
\usepackage{subcaption}     
\usepackage[normalem]{ulem}
\usepackage{array} 
\usepackage{enumitem}
\usepackage{colortbl}
\usepackage{xcolor}
\usepackage{framed}
\usepackage{fancyhdr}

\IEEEoverridecommandlockouts
\begin{document}
\title{Designing, Deployment and Field Testing of C2Stack for Networked Intelligent Software-Defined UAVs}

\author{Maxwell McManus$^1$,
Zhaoxi Zhang$^{2,3}$,
Sidharth Santhi Nivas$^{2,3}$, 
Yuqing Cui$^{2,3}$, \\
Prem Sagar Pattanshetty Vasanth Kumar$^1$,  
Chenzhi Zhao$^{2,3}$, 
Nicholas Mastronarde$^1$, \\
George Sklivanitis$^4$,  
Dimitris A. Pados$^4$,  
Elizabeth Serena Bentley$^5$,
and Zhangyu Guan$^{2,3}$\\
$^1$Department of Electrical Engineering, University at Buffalo, NY 14260, USA \\
$^2$Bradley Department of Electrical and Computer Engineering, Virginia Tech, VA, USA\\
$^3$Department of Computer Science \& Engineering, University of Minnesota - Twin Cities, MN 55455, USA\\ 
$^4$Department of Electrical Engineering and Computer Science, Florida Atlantic University, FL 33431, USA\\
$^5$U.S. Air Force Research Laboratory (AFRL), NY 13441, USA \\
Email: 
\{premsaga, memcmanu, nmastron\}@buffalo.edu,
\{joshzx, sidharths, yuqing, \\
chenzhiz, zhangyuguan\}@vt.edu,  
 \{gsklivanitis, dpados\}@fau.edu, 
elizabeth.bentley.3@us.af.mil 
\thanks{This work was supported in part by the National Science Foundation (NSF) under Grant SWIFT-2229563 and CNS-2450418, and the U.S. Air Force Research Laboratory under Contracts FA8750-21-F-1012, FA8750-20-C1021 and FA8750-25-1-1000.}
\thanks{Distribution A. Approved for public release: Distribution Unlimited: AFRL-2025-5623 on 11 Dec 2025.}
}

\maketitle

\fancyhead[L,R]{}
\fancyhead[C]{\small This work has been accepted for publication at ACM MobiCom 2026.}
\pagestyle{fancy} 
\thispagestyle{fancy} 
\renewcommand{\headrulewidth}{0pt}

\begin{abstract}
    Unmanned Aerial Vehicles (UAVs) are emerging as critical enablers of next-generation wireless networking and autonomous systems. Despite their potential, deploying and testing networked UAV systems in real-world environments remains challenging, largely due to the absence of well-developed, end-to-end, ready-to-use protocol stacks. To fill this gap, we present \textit{C2Stack}, a configurable protocol stack and experimental framework designed for real-time control, evaluation, and optimization of UAV networks. C2Stack incorporates a modular control plane, referred to as the~C2Stack Network Operating System (CNOS), alongside a programmable data plane that exposes APIs for cross-layer algorithm development, digital twin integration, and autonomous swarm control. 

    In this article, we share our experience with the deployment and testing of C2Stack. We implemented C2Stack on a custom UAV swarm platform that integrates multiprocessor system-on-chip (MPSoC) radios with Intel NUC computing modules, enabling interoperability with various RF front ends. Field trials were conducted in both netted environments and large-scale outdoor test ranges, focusing on two representative use cases: (i) network utility maximization through online reinforcement learning, and (ii) collaborative interference source localization. The experiments demonstrate the feasibility of real-time, data-driven optimization in dynamic aerial environments, while also revealing practical challenges in field deployments of networked UAV systems, including power constraints, sensing limitations, and deployment logistics. 
    We have made C2Stack source code available to the community under the MIT License\footnote{\href{https://wingslabwireless.github.io/opensource.html}{https://wingslabwireless.github.io/opensource.html}}, with the goal of establishing it as a foundational framework for experimental research on intelligent networked aerial systems.
\end{abstract}




\vspace{-4mm}
\section{Introduction}\label{sec:intro}

Unmanned Aerial Vehicles (UAVs) have emerged as a promising enabling technology for a wide range of applications, including urban air mobility (UAM) \cite{park2023MARLcooperativeUAM}, precision agriculture \cite{liu2024agriculture}, and environmental monitoring \cite{fairman2024waterproofuav}. They have also been applied to collaborative search and rescue \cite{abdellatif2025uavsearch}, on-demand network infrastructure deployment \cite{indu2024disasteruav, moorthy2022FlyTera, hu2021swarmshare}, distributed simultaneous localization and mapping (SLAM)~\cite{chang2025lidaruav}, autonomous supply chain optimization \cite{jahani2025supplychain}, distributed IoT network management \cite{jia2024ccdsUAV}, and large-scale additive manufacturing \cite{wang2024uavadditive}, as well as tactical operations. More recently, significant attention has been devoted to next generation (NextG) UAV-enabled networks, with use cases spanning mobile edge computing \cite{xu2024blockchainedge}, multi-modal network deployments~\cite{hu2025multimodal}, and adaptive traffic offloading~\cite{guo2025trafficdemand}.

Despite their great potential, the deployment and testing of networked UAV systems especially swarm UAVs in real world environments remains a highly challenging task. In particular, aerial operation for multirotor platforms induces vibrations across the airframe, leading to performance fluctuations in onboard wireless communication devices \cite{qi2024vibrationmodel}. Wind currents further exacerbate instability, introducing high variance in flight dynamics and reducing the positional reliability of hovering nodes and UAVs in motion \cite{dou2024wobble}. The combined effects of wind and vibration on UAV stability and device reliability are influenced by numerous factors, including rotor configuration, hardware controllers, calibration thresholds, processing capacity, and airframe design \cite{schweiger2023evtolconditions}. These physical instabilities not only degrade system robustness, but also increase the complexity of wireless channels, thereby affecting network performance in both air-to-air (A2A) and air-to-ground (A2G) links \cite{qian2025wobble, hua2025posturemodel}.

\begin{figure*}[t]
    \centering
    \includegraphics[width=0.84\linewidth,height=0.53\linewidth]{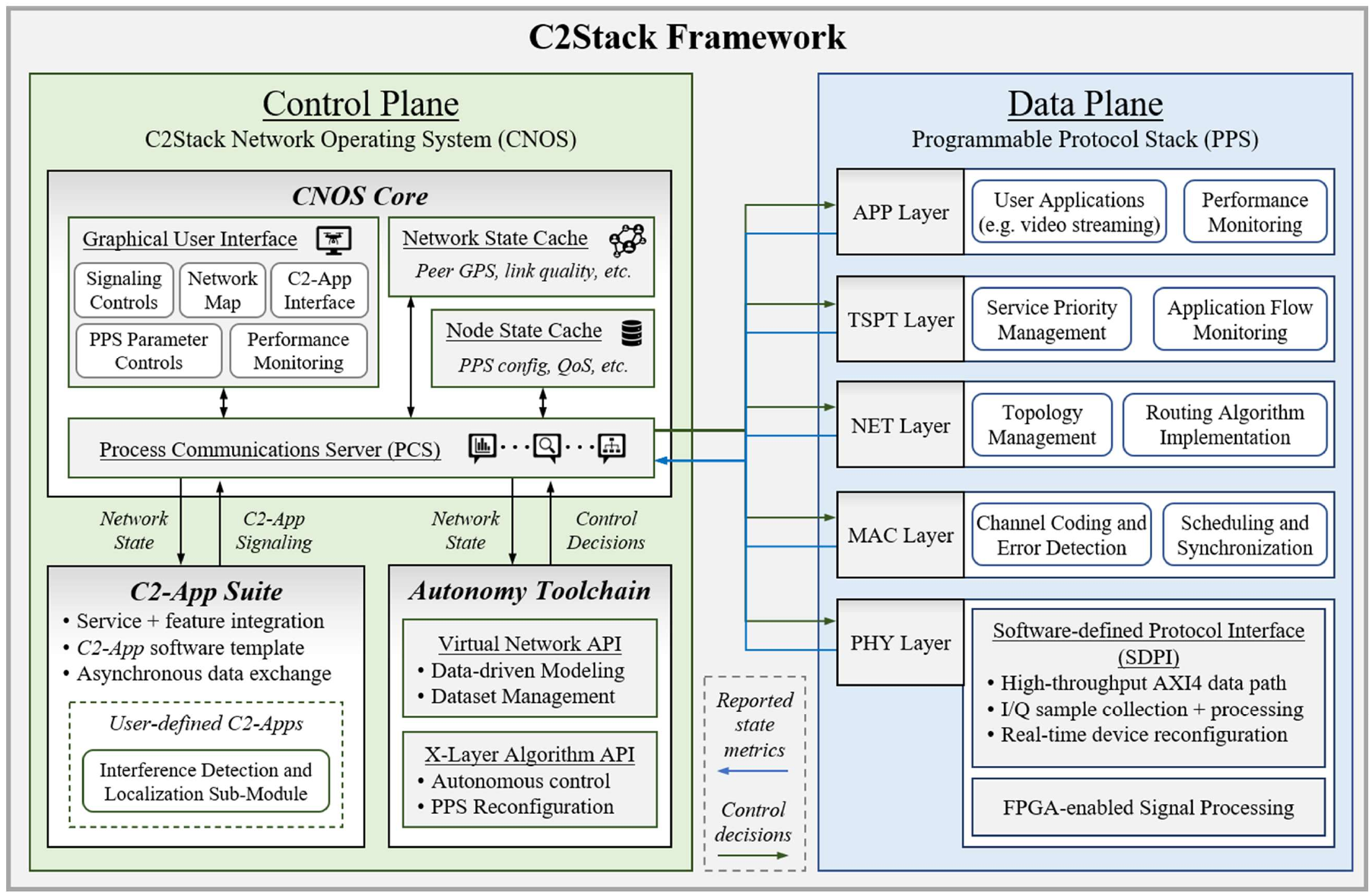}
    \vspace{-1mm}
    \caption{\small The \textit{C2Stack} framework: a configurable and extensible protocol stack for UAV-enabled wireless networks, supporting modular design, cross-layer interactions, and real-time control for autonomous swarm operation.}
    \vspace{-2mm}
    \label{fig:c2stack}
\end{figure*}

\textbf{Related Work and Limitations.}
There is a large and growing body of research in the existing literature that addresses the challenges of deploying, operating and controlling UAV networks \cite{sun2024genaiuav, phadke2024usmart, jia2024ccdsUAV, park2023MARLcooperativeUAM, schweiger2023evtolconditions, yin2024drlfixedwing, wu2024hybriddelay, ju2024multiuav, sehito2024nomauav, sabarish2022flytera, mcmanus2023survey, wang2024securitysurvey, javed2024uavresearchsurvey, dey2025airtwinx, moorthy2022simsocket, SwarmControlInfocom20,GuanToN20Drone,FlyBeamComNet2023}. 
For example, the U-SMART framework \cite{phadke2024usmart} improves the resilience of data-driven swarm control by modularizing key functions such as collision avoidance, energy modeling, and task allocation. Its centralized approach reduces overall complexity through periodic exchanges between UAVs and a central controller. Similarly, the cooperative cognitive dynamic system (CCDS) \cite{jia2024ccdsUAV} supports both centralized and distributed swarm management through separate modules for attention perception, learning and inference, and risk control. By decomposing swarm decision-making into independent sub-problems, CCDS improves computational efficiency.

Despite these advances, a major limitation of most existing efforts, particularly those leveraging AI/ML models and control algorithms, is their reliance on idealized network conditions, simplified communication models, or highly customized datasets \cite{feng2024nomauav, cui2024byzantineattack, taye2023UAMtrajectory, park2023UAMSpaceAirGround, cattai2025uavsurvey}. Although existing frameworks show promising performance, their effectiveness in real networks cannot be accurately assessed without extensive field trials using over-the-air (OTA) wireless devices. However, establishing a reliable workflow for OTA validation remains a significant barrier, as it requires addressing both the inherent instability of UAV platforms and the complexity of real-world wireless environments. To the best of our knowledge, there is still no well-developed end-to-end protocol stack readily available to support rigorous and systematic research on UAV networking.

Access to hardware and facilities for UAV-enabled experiments presents another significant research barrier. We have identified several public testbed platforms which seek to enable practical research in this domain: AERPAW \cite{marojevic2020aerpaw}, and ARA \cite{zhang22ara}. AERPAW specializes in UAV-enabled research, and provides both virtual and OTA experimental resources as well as consultation services. However, the scope of experiments supported by AERPAW is narrow, due to low UAV payload capacities, lack of support for swarm deployments, and limited communication protocol options. The platform imposes additional restrictions on flight control strategies, user software integration, and experiment scheduling, further reducing experiment flexibility. In contrast, while ARA has been used for UAV-enabled research \cite{babu2025ara_uav}, use cases are not clearly documented, suggesting that this is outside of the normal scope of operation and unavailable for general UAV research. Additionally, different from existing initiatives, C2Stack aims to provide a toolchain to accelerate the development and evaluation of UAV-enabled network protocols, and hence expand the scope of supported research on existing testbeds through collaborative development.


\textbf{Contributions.}
To bridge the gap between theoretical and experimental research in networked UAV systems, we present the design, implementation and evaluation of \textit{C2Stack}, a configurable and extensible \textit{protocol stack for communications and control} in UAV networks. C2Stack supports real-time user interaction with deployed UAVs during aerial operations and provides signaling interfaces between UAVs to enable autonomous swarm control and wireless network self-configuration through data-driven control algorithms.

The primary contributions of this work are as follows:

\begin{itemize}[leftmargin=12pt]
\item \textit{C2Stack Design}.  
We present \textit{C2Stack}, a fully configurable protocol stack for the rapid deployment and evaluation of advanced modeling and control techniques in UAV-enabled wireless networks. The stack is designed to support the integration and testing of AI/ML algorithms in real-world scenarios, emphasizing dynamic end-to-end network control and cross-domain interactions. Each C2Stack protocol layer is implemented in a modular fashion to maximize adaptability, while well-defined APIs enable user-defined algorithms to interact directly with core system components. 

\item \textit{Aerial Platform Development}.  
We design and implement an aerial platform to facilitate experimental research on the capabilities and limitations of networked UAV systems. A fleet of custom medium-duty UAVs is built to carry \textit{C2Stack} payload modules. The platform incorporates a configurable interface that enables real-time monitoring and control of bidirectional data streams between field-programmable gate array (FPGA)-based baseband signal processing and the upper protocol layers.

\item \textit{Field Deployment and Testing}.  
We deploy and operate the UAV network developed to demonstrate the capabilities of \textit{C2Stack} for practical evaluation of data-driven network modeling and control. We focus on two representative control problems, dynamic protocol self-configuration and interference source estimation, and present benchmark results from real-world deployments. In addition, we share insights gained during the \textit{C2Stack} development process to highlight the complexity of implementing UAV-enabled wireless systems, identify limitations of the initial swarm implementation, and outline key research challenges that define a roadmap for future work.

\item \textit{Community Release}. 
We will release the C2Stack framework on GitHub under the MIT open-source license \cite{mit_license} for community use in a timely manner. A dedicated GitHub repository will be created to host the C2Stack source code and associated materials. The release will include a user manual outlining installation, general operation, code integration, and the steps required to replicate the experiments presented in this paper. In addition, we will provide detailed specifications of the UAV platform, payload design, and auxiliary equipment used in the experiments.
\end{itemize}

\noindent The remainder of this paper is organized as follows. Section~\ref{sec:framework} introduces the proposed \textit{C2Stack} architecture, and Section~\ref{sec:swarm_development} describes the development of the aerial platform. The experimental demonstration is presented in Section~\ref{sec:mission}, followed by results and lessons learned in Sections~\ref{sec:results} and~\ref{sec:lessons}, respectively. Finally, we draw the main conclusions in Section~\ref{sec:conclusion}.

\section{C2Stack Framework Design} \label{sec:framework}

The objectives of the C2Stack design are twofold: to enable real-time configuration and evaluation of various network protocols, data-driven control algorithms, and adaptive modeling techniques; and to provide experimenters with fine-grained control over individual processes of the protocol stack, 
from physical-layer waveforms to custom network applications, 
with an emphasis on adaptive cross-layer control. In this way, C2Stack not only allows theoretical designs to progress beyond simulation and small-scale experimentation to rapid validation in real-world deployments, but also facilitates systematic investigation of how novel modeling and control algorithms interact with established approaches.

Toward this end, as illustrated in Fig.~\ref{fig:c2stack}, we focus the C2Stack design on two core elements: a control plane and a data plane. The control plane, referred to as the \textit{C2Stack Network Operating System} (CNOS), enables monitoring, visualization, and adaptive management at both the node and network levels for C2Stack-enabled devices. The data plane, built around a \textit{Programmable Protocol Stack} (PPS), serves as the primary communication framework by providing a software-defined implementation of layer-specific communication protocols aligned with the traditional OSI model. The interaction between the control and data planes affords fine-grained control over the communication configuration and decision-making processes.

\subsection{C2Stack Control Plane}\label{sec:c2stack_cnos}
The CNOS provides coordinated management of both node-level and network-level processes. At the node level, CNOS coordinates endogenous functions such as protocol adaptation, local environment virtualization, and real-time execution of control decisions. At the network level, it supports exogenous functions including topology reconfiguration, collaborative situational awareness, and end-to-end swarm optimization. CNOS is organized into three primary modules: the \textit{CNOS Core}, which provides the fundamental runtime and system services; the \textit{Autonomy Toolchain}, which enables adaptive decision making and control; and the \textit{C2-App Suite}, which provides user-facing applications for customized network experimentation.

\textbf{CNOS Core}. The CNOS Core provides the interface for the management, control, and monitoring of node- and network-level processes. At its center is a \textit{process communication server} (PCS), which coordinates all inter-process exchanges among local CNOS services, facilitates PPS signaling, and manages CNOS signaling with neighbor nodes through the C2Stack control path.  To ensure seamless coordination, CNOS employs a unified message protocol that routes messages across the PCS through three logical interfaces:  i) \textit{internal}, handling exchanges among processes and modules within CNOS;  
ii) \textit{node-level}, coordinating exchanges between CNOS and the PPS; and iii) \textit{network-level}, supporting signaling among CNOS instances across different nodes. These interfaces are implemented through an encapsulation procedure that defines standard headers, complemented by extensible sub-headers designed to reduce message complexity. Two sub-header sets have been designed: a \textit{control plane} set, which establishes signaling routes over the internal and network-level interfaces; and a \textit{data plane} set, which establishes signaling routes over the node-level interface.  

\begin{figure}[t]
  \centering
  \vspace{-8mm}
  \includegraphics[width=1\linewidth]{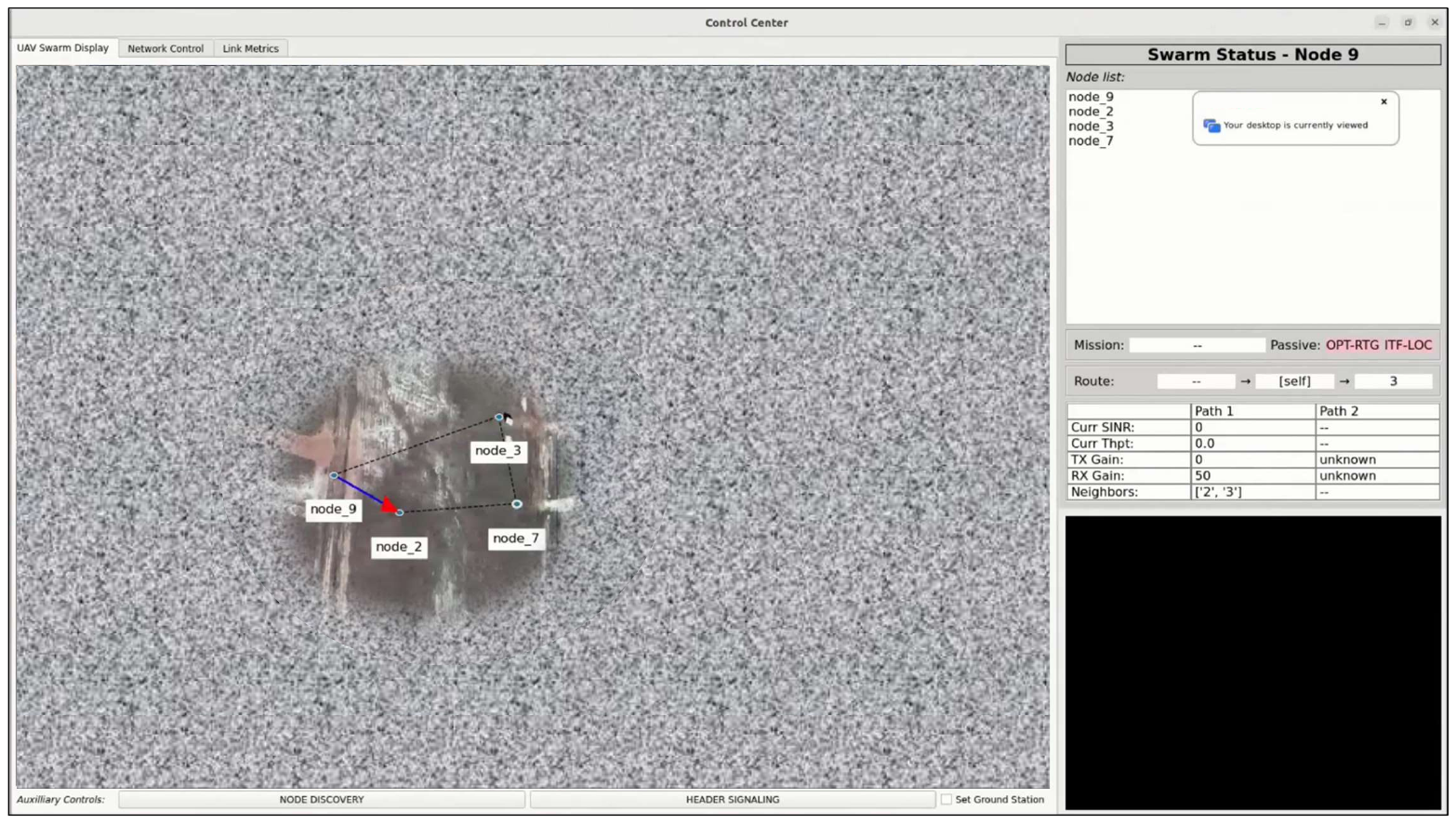}
  \vspace{-16mm}
  \caption{\small CNOS GUI live network map display, showing real-time PPS state metrics and rendering the deployment area using GPS updates. Note: The deployment area is censored to preserve anonymity.}
  \label{fig:cnos_gui_map}
  \vspace{-4mm}
\end{figure}

\begin{figure}[ht]
  \centering
  \vspace{-8mm}
  \includegraphics[width=1\linewidth]{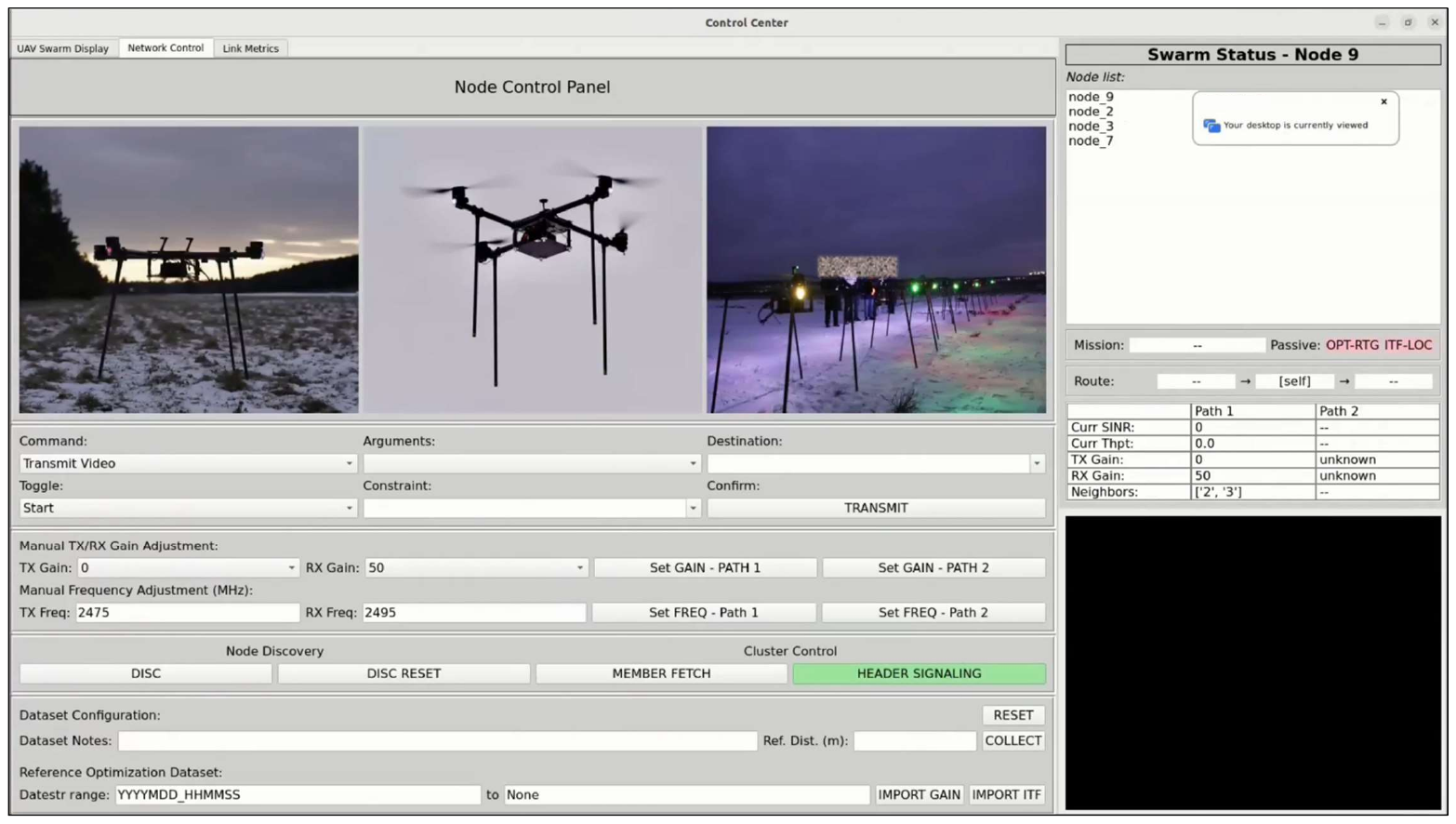}
  \vspace{-16mm}
  \caption{\small CNOS GUI control panel, allowing users to adjust PPS parameters during operation and interface with both the Autonomy Toolchain and C2-Apps through the backend.}
  \label{fig:cnos_gui_ctrl}
  \vspace{-4mm}

\end{figure}

\begin{figure}[ht]
  \centering
  \vspace{-8mm}
  \includegraphics[width=1\linewidth]{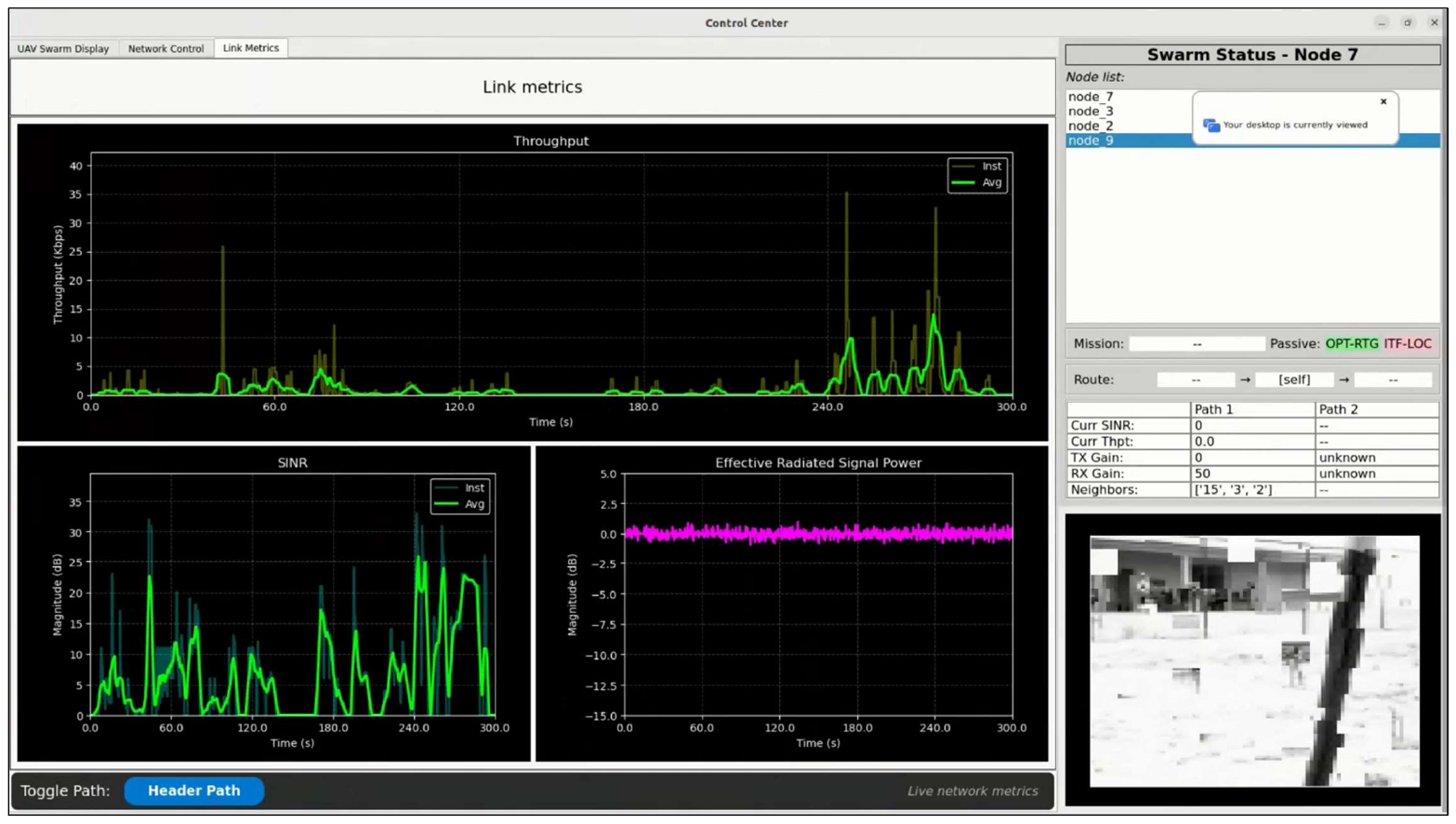}
  \vspace{-16mm}
  \caption{\small CNOS GUI metric display, plotting throughput, SINR, and transmit power over time, with an embedded active video stream in the bottom-right window.}
  \label{fig:cnos_gui_metric}
  \vspace{-4mm}
\end{figure}

Additionally, leveraging the PCS interface, the CNOS GUI enables real-time user control over APP-layer services, node-level protocol configurations, and network-level optimization strategies, thereby supporting dynamic, intent-driven operation of the UAV network. A snapshot of the GUI is shown in Figs.~\ref{fig:cnos_gui_map}-\ref{fig:cnos_gui_metric} on the next page. Specifically, Fig.~\ref{fig:cnos_gui_map} illustrates real-time network deployment, including PPS state metrics such as throughput, routing decisions, and neighbor discovery, while also rendering the deployment area using satellite tiles via the Cartopy library \cite{Cartopy}. Fig.~\ref{fig:cnos_gui_ctrl} presents the GUI control panel, which allows users to adjust the PPS parameters during operation and to interface with the Autonomy Toolchain and C2-Apps through the GUI backend. Finally, Fig.~\ref{fig:cnos_gui_metric} plots performance metrics over time, such as throughput, SINR, and transmit power, and displays the active video stream in the bottom-right window.

\textbf{Autonomy Toolchain}. The toolchain is designed to enable autonomous UAV network control by providing a set of APIs for implementing user-defined optimization algorithms with the node- and network-level state information available in the CNOS Core module. The toolchain operates synchronously with the CNOS Core to adapt in real-time to observations from the PPS. Two key APIs have been developed to facilitate this: the \textit{X-Layer Algorithm API} and the \textit{Virtual Network API}. The former consolidates the integration workflow by providing a list of available metrics to fetch from the CNOS Core, default protocol parameters available for resource orchestration, and templates for defining custom control specifications. The latter provides a data publication interface through which the same metrics and parameters can be exchanged between a virtual networking environment and CNOS Core services. The Autonomy Toolchain also supports importing stored datasets during network runtime for dynamic initialization of offline learning algorithms to accelerate prototyping and evaluation of a wide range of network control algorithms. Both APIs leverage the PCS to support interaction with the CNOS GUI and the PPS.

\textbf{C2-App Suite}. To extend support for various modeling and control missions, the C2-App Suite provides an asynchronous interface that allows user-defined modules and additional feature sets to be integrated with existing control plane services via the PCS. This design enables C2Stack to accommodate a wide range of experimental applications at both the node and network levels, without requiring extensive reconfiguration of existing toolchains.  
As an illustrative example, we developed the \textit{Interference Detection and Localization Sub-Module} (IDLS) within the C2-App framework. Operating as an independent program, the IDLS can request state data from the CNOS Core, such as node and neighbor locations, network metrics, and PHY-layer configurations, and forward relevant results to the CNOS GUI for real-time visualization.

\vspace{-3mm}
\subsection{C2Stack Data Plane}\label{sec:c2stack_pps}

At the core of the C2Stack data plane is the PPS, which coordinates different protocol layers for data forwarding and enables dynamic cross-layer optimization. A design objective of the PPS is to allow characteristic processes at each layer, e.g., network-level routing or MAC-layer scheduling, to be deployed synchronously during network operation. We note that the focus of this work is not on developing new protocols for individual layers. Instead, the PPS is designed as a programmable suite of independent protocol processes that serve as a platform to demonstrate how CNOS enables real-time monitoring and control of network behavior.

\subsubsection{PPS Architecture} 
To maximize compatibility with existing protocol stacks, the PPS is designed following a five-layer architecture: the application layer (APP), the transport layer (TSPT), the network layer (NET), the medium access control layer (MAC), and the physical layer (PHY). The APP layer provides the primary data generation service within C2Stack. It hosts user-defined applications and supports adaptive data generation.
The TSPT layer provides priority-based flow management across APP-layer services, and enables flexible reliability guarantees via end-to-end packet-level acknowledgment and retransmission. The NET layer maintains topology information and manages both node discovery and routing. The discovery process establishes associations among active nodes based on user input, node role assignments, or RF front-end configuration. The routing process determines packet reception and forwarding decisions using user-defined rules and PPS metrics. The MAC layer supports error detection and correction mechanisms, such as channel coding and cyclic redundancy checks (CRC). It also performs queue monitoring and tracks both node-specific and aggregate throughput, enabling scheduling algorithm integration.  

Finally, as shown in Fig.~\ref{fig:c2stack}, the PHY layer includes both the connected RF hardware and a software-defined protocol interface (SDPI). The SDPI manages data transfer between software and hardware processes, including payload aggregation, data type conversion, bit masking, and symbol mapping. It also monitors waveform-level metrics such as access code symbol error rate, error vector magnitude (EVM), and channel distribution estimates.

\subsubsection{Layer Development of PPS}\label{sec:pps_config}
To support the testing campaign, we implement PPS with an interference-resistant wideband protocol for A2A and A2G links. Although this protocol is used as an example, the C2Stack framework is interoperable with any open-source ad hoc networking protocols through adaptation or replacement of the relevant layers. The modularity of C2Stack is achieved through the independent, object-oriented implementation of each protocol layer and the abstraction of their associated services. For example, routing at the NET layer and scheduling at the MAC layer are implemented as independent objects within their respective execution loops. This allows user-defined algorithms to extend or replace existing services with minimal code-level modification. Moreover, services at each layer can interact with co-located processes via the stateful layer object, or exchange information asynchronously with other layers using built-in packet types and signaling pathways.

\textbf{APP layer}. The APP layer hosts a distributed surveillance application in which a video feed is relayed through a multi-hop UAV network from a source node to a destination. Each video frame is sliced, compressed via discrete cosine transform \cite{thota2008dct}, packetized, and forwarded to the TSPT layer for transmission. 
A buffer-less approach is employed at the receiver: video packets are received, decompressed, and displayed in order of arrival without full-frame reconstruction.

\textbf{TSPT layer}. The TSPT layer implements a generalized flow control algorithm to manage the delivery of APP-layer packets. Independent flow identifiers are assigned to each pair of source and destination UAVs, enabling queue management along the APP data path. 

\textbf{NET layer}. The NET layer supports discovery of nodes that share frequency resources, enabling manual configuration of network topology at runtime. It also implements a lightweight link-state routing algorithm, inspired by the OLSR protocol, which leverages CNOS feedback to optimize next-hop selection according to network performance.

\textbf{MAC layer}. The MAC layer manages the data and control path queue processes, packet-level throughput measurement, as well as packet quality assessment via CRC to detect and discard corrupted packets along the receive path.

\textbf{PHY layer}. Because UAV systems often operate in scenarios where resilient communication is critical, the PHY layer employs a direct sequence spread spectrum (DSSS) waveform implemented on FPGA fabric to accelerate baseband processing compared to software-based approaches. The SDPI enables real-time adaptation of key parameters, including frequency, modulation, front-end gains, and code selection. These parameters can be accessed via Autonomy Toolchain, C2-Apps, or GUI, through the CNOS PCS, supporting configurable adaptation of transmit and receive processes.

\begin{figure}
    \centering
    \includegraphics[width=0.99\linewidth]{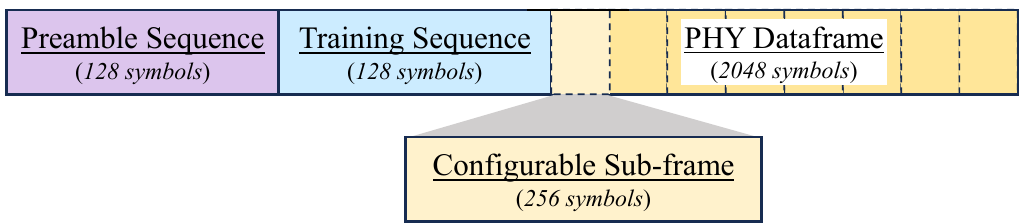}
    \caption{\small C2 PHY-layer frame structure consisting of preamble, training sequence, and dataframe divided into eight configurable subframes.}
    \vspace{-4mm}
    \label{fig:frame_structure}
\end{figure}

\subsubsection{C2 Frame Structure}\label{sec:pps_frame}

At the PHY layer, the overall design of the frame is shown in Fig.~\ref{fig:frame_structure}. The current waveform uses a frame size of 2304 symbols, consisting of three main components: (i) a 128-symbol preamble sequence for frame detection, (ii) a 128-symbol training sequence for coarse frequency synchronization, and (iii) a 2048-symbol PHY-layer dataframe, which is logically divided into eight configurable subframes of 256 symbols each. Depending on modulation and channel conditions, each subframe can be configured via the SDPI to include between one and eight 16-symbol pilot sequences, used for for fine frequency synchronization and phase offset correction, with remaining symbols allocated for data. Since this division is determined before forwarding the dataframe to the FPGA, it can be easily adapted or removed to accommodate different packet size requirements.


\section{Aerial Platform Development}\label{sec:swarm_development}

To test and deploy the proposed framework and to leverage the configurable nature of C2Stack, we developed a large-scale aerial experimentation network with custom medium-duty UAVs. According to both the existing literature \cite{dou2024wobble, qi2024vibrationmodel} and our prior experience, we determined that the maximum payload capacity for each single UAV should exceed the target payload weight by a substantial margin (approximately 150\%). In the following, we describe the UAV platform and the C2Stack payload design in sequence.


\subsection{UAV Platform Design}\label{sec:uav_design}

\begin{figure} [t]
    \centering
    \includegraphics[width=0.98\linewidth]{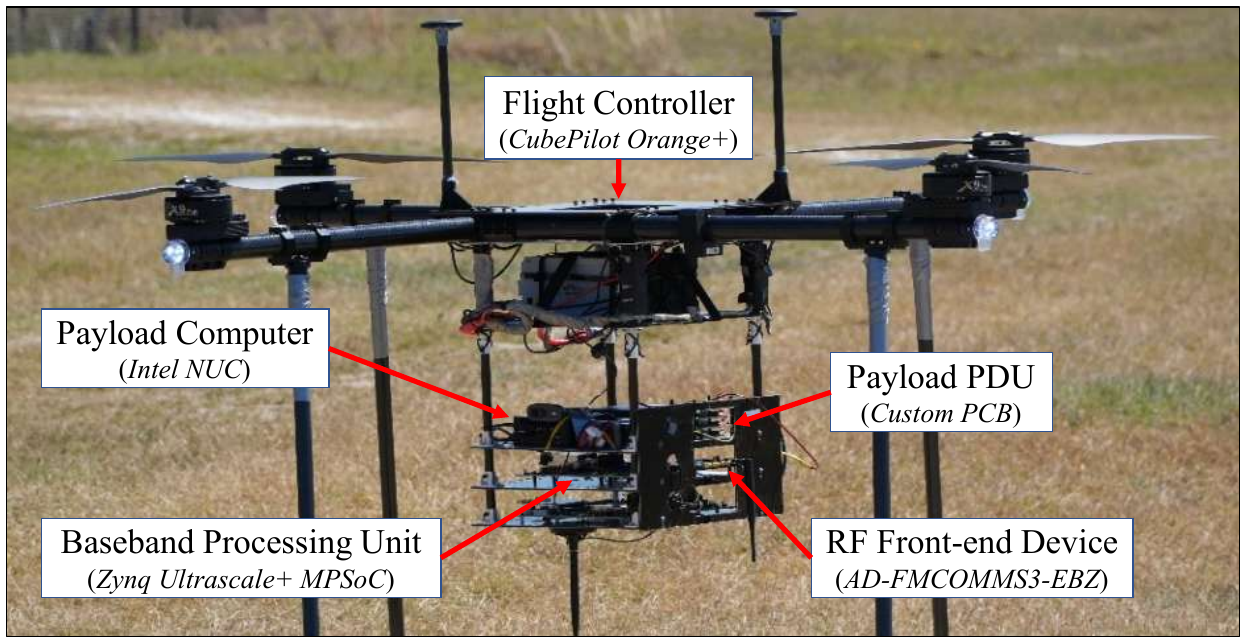}
    \caption{\small C2Stack prototype payload module deployed on custom medium-duty UAV platform.}
    \vspace{-4mm}
    \label{fig:drone_payload}
\end{figure}

The UAV platform was developed as a medium-duty quadcopter to provide stable flight, high payload capacity, and modularity for networking experiments. As shown in Fig.~\ref{fig:drone_payload}, the body design consolidates rotor placement to enhance stability and reduce vibration during operation. The primary airframe is constructed from carbon fiber, selected for its high strength-to-weight ratio and vibration resistance, with auxiliary connectors fabricated using 3D-printed carbon-fiber PLA. The flight controller is centrally mounted on an aluminum bracket to minimize vibration effects, while carbon fiber landing legs replace traditional skids to reduce weight. Aluminum payload couplers allow for rapid installation and exchange of experimental payloads. The power of the UAV is supplied by 12S–14S smart LiPo batteries with integrated battery management, mounted on the airframe underbelly for optimal balance and accessibility. For safety and operational reliability, UAV flight is controlled using a long-range radio controller operating in the 5.8 to 6 GHz band. The MAVLink protocol \cite{mavlink} is employed for telemetry and command exchange, ensuring robust and extensible communication between the UAV and the operator. A swarm of these experimental UAVs was constructed to demonstrate the capabilities of C2Stack. Although this custom platform is used as the reference design in our work, the modular nature of C2Stack allows deployment on any UAV platform capable of hosting standalone computing and RF hardware. 

We note that the C2Stack hardware and structural payload components weight roughly 3.17 kg and the custom PDU adds 0.91 kg, resulting in a total payload weight of 4.08 kg. Considering 20\% payload redundancy to improve flight stability, this module could be adapted for use on any UAV platform with a capacity $\geq$5 kg, such as the Freefly Alta X \cite{freefly_alta_x} or the Aurelia X6 \cite{aurelia_x6}. Deployment on light-duty UAV platforms, such as the DJI Matrice 350 RTK \cite{dji_matrice_350rtk}, could be achieved by replacing FPGA and compute hardware with lighter counterparts and reducing the payload form factor, which would introduce significant tradeoffs in experimental flexibility, computational capacity, and flight duration.

\subsection{C2Stack Payload Implementation}\label{sec:c2stack_module}
The C2Stack protocol is installed on the payload computer, which is an Intel NUC. The NUC hosts the control plane and upper layers of the PPS and is interfaced with a Zynq UltraScale+ ZCU102 MPSoC platform for baseband signal processing. The MPSoC provides a reconfigurable FPGA fabric for high-performance transmit and receive operations, while the NUC delivers general-purpose computing capability in a compact form factor. As shown in Fig.~\ref{fig:drone_payload}, an AD-FMCOMMS3-EBZ front-end device integrated with the MPSoC enables transmission and reception over a 70~MHz–6~GHz frequency range with up to 56~MHz bandwidth.  

To enhance generality and evaluate the flexibility of configurable radio devices, only the transmit and receive chains specific to the PHY-layer protocol described in Sec.~\ref{sec:c2stack_pps} are implemented on the MPSoC, while all other protocol functions are handled by the SDPI module on the NUC. This design ensures the interoperability of the C2Stack framework with a variety of communication devices, including software-defined radios such as Ettus USRP platforms \cite{Ettus} and other FPGA-enabled devices such as RFSoC \cite{rfsoc}.

The prototype hardware is powered by an independent supply through a custom designed power delivery unit (PDU), which isolates the C2Stack payload from the UAV onboard power system to prevent resource contention. As shown in Fig.~\ref{fig:drone_payload}, the complete C2Stack module is mounted on the UAV undercarriage for deployment.

\section{Control Mission Design} \label{sec:mission}

The objectives of this initial experimental campaign are twofold. First, to demonstrate the potential of C2Stack as a platform for AI/ML-driven experiments across a range of network configurations. Second, to evaluate the practical experience, challenges, and limitations of deploying data-driven algorithms in UAV networking experiments, thereby providing insights to guide future efforts toward improving the technical readiness of UAV-enabled networking platforms.  

To this end, we investigate two representative network control missions to assess the field performance of C2Stack: (i) reinforcement learning (RL)-based link-level QoS optimization, and (ii) collaborative interference source localization. These missions represent two critical aspects of UAV swarm operation. The first aims to reduce user involvement during network startup by autonomously adapting control parameters throughout operation, accelerating deployment and improving performance under time-varying conditions. The second focuses on sustaining service quality during navigation while mitigating the impact of swarm spectrum usage on co-located or incumbent wireless networks. 

\subsection{Mission 1: Link-level QoS Optimization}\label{sec:protocol_optimization}

In UAV networks, onboard RF hardware (i.e., the payload described in Section~\ref{sec:c2stack_module}) must adapt to dynamic environments that cannot be fully characterized prior to deployment. During flight, a static protocol configuration may lead to suboptimal performance, necessitating real-time parameter adaptation. This mission employs C2Stack for protocol self-optimization, dynamically reconfiguring RF front-end parameters to enhance network QoS throughout operation.

\textbf{Problem Statement.} We formulate the link parameter optimization task as a $K$-armed bandit problem.
Consider a swarm $\mathcal{U}$ of $N_{U}$ connected UAV nodes, defined as $\mathcal{U}=\{U_{m}\}, m=1,2,\ldots,N_{U}$. An independent RL agent is deployed at each $U_{m}\in\mathcal{U}$, with the objective of maximizing the aggregate link throughput
$\Bar{R}_{m}(\tau) = \sum_{m \neq n} R_{m,n}(\tau)$, where $R_{m,n}(\tau)$ denotes the average throughput of the link originating from $U_{n}$ observed at $U_{m}$ during interval $\tau$.  

We consider that each RL agent is restricted from directly interacting with user-defined protocol processes or communicating with neighbor nodes. Consequently, the set of $K$ available actions for the agent in each interval $\tau$ corresponds to the discrete set of supported RF front-end gain values, denoted as $G_{rx} = \{g_{k}\}, k=1,\ldots,K$. In each interval $\tau$, an agent selects an action $a_{m}(\tau)\in G_{rx}$ and observes $\Bar{R}_{m}(\tau)$. 
The agent then receives a reward $r_{m}(\tau)\in \{-0.1,0,1\}$, defined as
\[
r_{m}(\tau) =
\begin{cases}
1, & \Bar{R}_{m}(\tau) \geq \delta \cdot \max_{t<\tau}\Bar{R}_{m}(t), \\[1pt]
-0.1, & \Bar{R}_{m}(\tau) < (1-\delta)\cdot \max_{t<\tau}\Bar{R}_{m}(t), \\[1pt]
0, & \text{otherwise},
\end{cases} 
\]
where $t\in \mathbb{Z}^{+}$. To account for variance in $\Bar{R}_{m}(\tau)$, we set $\delta=0.9$.  
The probability of selecting action $a_{m}(\tau)=g_{k}$ in $\tau$ is given by $P_{k}^{m}(\tau)$, where each $g_{k}\in G_{rx}$ is assigned a weight $W_{k}^{m}(\tau)$ to estimate $P_{k}^{m}(\tau)$. The objective of the agent at node $U_{m}$ is to maximize $\Bar{R}_{m}(\tau)$ over the flight duration $T_{m}$ by adaptively selecting the optimal $g_{k}\in G_{rx}$.

To solve this problem, we implement the lightweight MIX-MAB algorithm \cite{azizi2022mixmab} in the Autonomy Toolchain described in Sec.~\ref{sec:framework}. We note that the objective of this work is not to propose new machine learning algorithms, which we leave as a direction for future research.
The algorithm is implemented in two phases: the \textit{exploration} phase, and the \textit{continuing} phase. Initially, $P_{k}^{m}(0) = 0, \forall{m,k}$ and $W_{k}^{m}(0)=1, \forall{m,k}$. During each slot $\tau$ of the exploration phase, each RL agent selects $a_{m}(\tau)\in G_{rx}$ by round-robin selection and observes reward $r_{m}(\tau)$. 
First, the agent updates $P_{k}^{m}(\tau+1)$ using $P_{k}^{m}(\tau+1)= (1-\gamma) \cdot \frac{W_{k}^{m}(\tau)}{\sum_{k}W_{k}^{m}(\tau)} + \frac{\gamma}{K}$, 
where $\gamma$ is the learning rate of the RL agent. The value of $P_{k}^{m}(\tau+1)$ is then normalized across $G_{rx}$. 
Using the new next-step action probability $P_{k}^{m}(\tau+1)$, the weight of the selected $g_{k}$ in the next slot is updated using $W_{k}^{m}(\tau+1)=W_{k}^{m}(\tau) \cdot e^{\frac{\gamma \cdot r_{m}(\tau)}{K\cdot P_{k}^{m}(\tau + 1)}}$.
After each $g_{k}\in G_{rx}$ has been selected $N^{l}$ times, where $N^{l}$ is defined as the exploration threshold, the algorithm enters the \textit{continuing} phase. We set $N^{l}=1$ to minimize the time spent observing sub-optimal actions. Now, after $W_{k}^{m}(\tau+1)$ is updated, the set of probabilities is evaluated to identify sub-optimal actions. Specifically, considering an action removal threshold $\eta\in(0,1)$, if $P_{k}^{m}(\tau + 1) < \eta\cdot\max_{k}P_{k}^{m}(\tau)$, the action is considered sub-optimal and we can set $P_{k}^{m}(\tau + 1)=0$. Typically, $\eta=0.5$, although this can be increased to accelerate convergence or decreased to encourage exploration. 

The resulting action distribution is strictly reliant on the consistency of the deployment scenario across $\tau\in T_{m}$, and changes to network configuration or environmental dynamics will result in variance across $\Bar{R}_{m}(\tau)$ for each $g_{k}$ \cite{zhang2024cexp3}. However, resetting $P_{k}^{m}(\tau+1)=\frac{1}{K},\forall{k}$ upon detection of these changes allows RL agents to leverage prior experience via $W_{k}^{m}(\tau)$ to accelerate training in this new context.

\subsection{Mission 2: Interference Localization}\label{sec:interference_localization}
Whereas the first mission focuses on link-level optimization to adapt to large-scale variations in signal quality, certain deployment scenarios present more severe spectrum challenges that cannot be addressed through parameter tuning alone. In this mission, we demonstrate the ability of C2Stack to leverage observed network states for collaborative decision-making among UAV nodes, enabling interference-aware path planning at the flight controller without dedicated hardware or disruption to ongoing network operations.  

This functionality is provided by the IDLS described in Sec.~\ref{sec:c2stack_cnos}, which interprets fluctuations in runtime network metrics to generate interference-related insights for flight control. Since multiple nodes are assumed to share the given frequency band, interference source localization is achieved by aggregating predicted distances with the known positions of the detecting nodes, thereby enabling triangulation.

\begin{figure}[t]
    \centering

    \includegraphics[width=0.98\linewidth,height=0.4\linewidth]{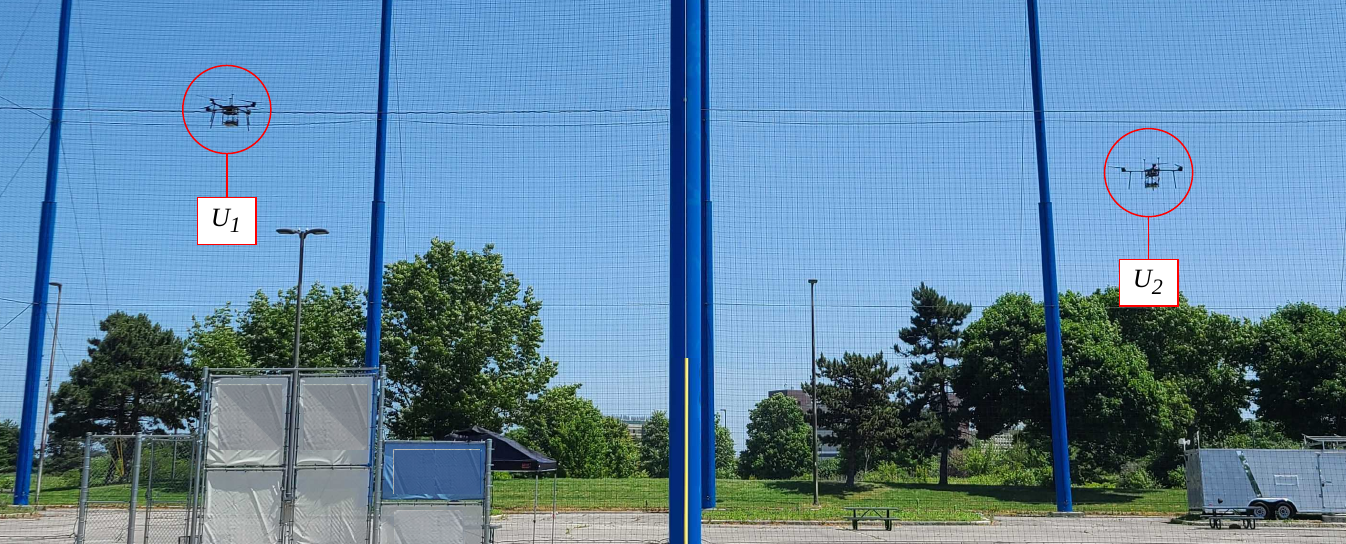}
    \caption{\small Flight experiment at netted UAV test facility showing operation of two UAV nodes.}
    \vspace{-3mm}
    \label{fig:soar_flight}
\end{figure}

\begin{figure}[t]
    \centering
    \includegraphics[width=0.98\linewidth]{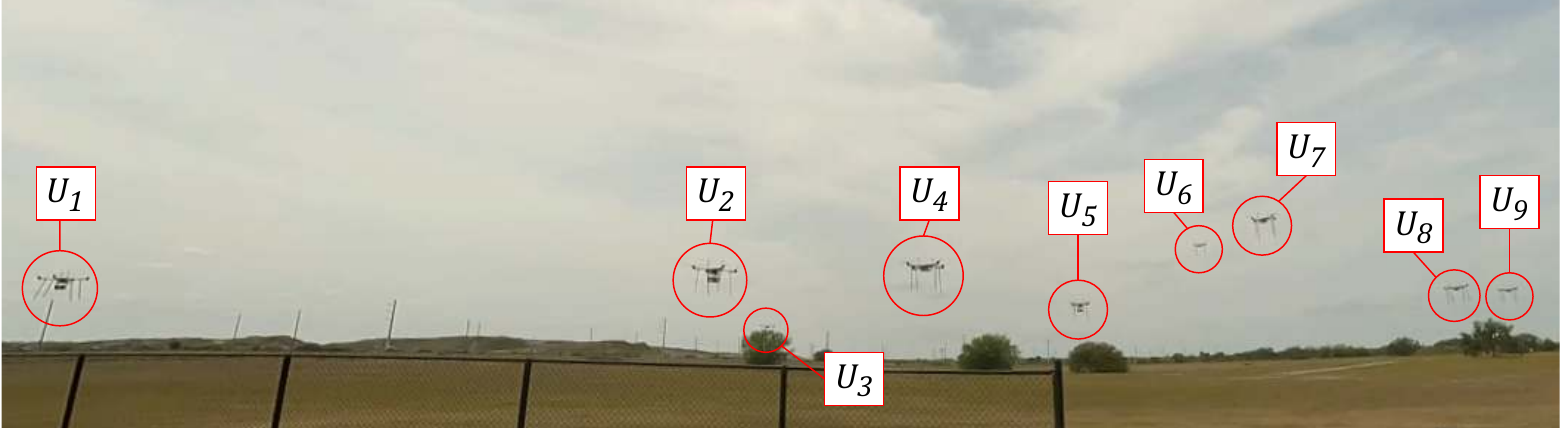}
    \caption{\small Flight experiment at open-air test facility showing operation of nine UAV nodes.}
    \vspace{-4mm}
    \label{fig:duette_swarm}
\end{figure}

\textbf{Problem Statement.}  
To predict and avoid locations of high interference, we develop a random forest regression model to estimate the distance between each deployed node $U_{m}\in\mathcal{U}$ and a local RF interference source $V$. Unlike traditional localization approaches that require dedicated RF sensing hardware or integrated sensing and communication (ISAC) capabilities \cite{peng2024irsisac, liu2023isac5g}, such as angle of arrival (AoA) \cite{zhang2018wsnaoa, yang2023wifiaoa} or time difference of arrival (TDoA) \cite{ting2019tdoa, kocher2024jamdetect}, in this mission we investigate a hardware-agnostic approach that does not rely on low-level signal analysis.

We consider a set of features $\Omega$ reported by the PPS, as described in Sec.~\ref{sec:pps_config}, which include the first- and second-order statistics of several network metrics: $\Bar{R}_{m}(\tau)$, pilot symbol error rate (SER), EVM, estimated SINR, and estimated channel distribution parameters. Training data are collected offline by varying the location of $V$ within the deployment area and associating entries in $\Omega$ with the ground-truth distance $d_{m}(\tau)$ between $V$ and $U_{m}$ at time slot $\tau$.  

To identify the most informative features, a Boruta feature selection process \cite{subbiah2022borutaforest} is applied prior to training. This method ranks feature importance by comparing each feature against shuffled copies of the dataset. Specifically: (1) copy all original features and shuffle the copies; (2) train a model on the extended dataset; (3) compute feature importance; (4) retain original features that rank higher than the most important copy; (5) repeat steps (1)–(4) for $N^{B}$ iterations to validate feature importance; and (6) extract $\Omega^{*}$, the subset of $\Omega$ containing the most relevant features.  

Finally, the random forest regression model is trained using an accelerated grid search procedure. Key hyperparameters, including the number of estimators, maximum tree depth, and maximum samples per leaf, are first coarsely tuned using randomized search, then fine-tuned through exhaustive grid search over the most promising ranges.

\vspace{-2mm}
\section{Field Measurements and Analysis}\label{sec:results}

In this section, we evaluate the C2Stack protocol introduced in Section~\ref{fig:c2stack} on the aerial platform described in Section~\ref{sec:framework}, focusing on the two missions outlined in Section~\ref{sec:mission}. We first describe the network environments and experimental methodology, then analyze the measurement results.

\vspace{-2mm}
\subsection{Network Environments}\label{sec:deployment_environments}

The two missions described in Section~\ref{sec:mission} were evaluated in two environments:  
(i) a dedicated UAV test facility with a netted outdoor enclosure for small-scale flight trials, and  
(ii) a large multi-use test facility offering a ground-accessible deployment area of approximately 200 acres, along with access to 21,000 acres of adjoining airspace.  
The netted enclosure is exempt from regulations governing medium-duty UAV operations, whereas for the open-air facility, licenses were obtained for all UAVs in accordance with FAA requirements.

\textbf{Netted Enclosure Deployment.}  The small-scale network deployment is illustrated in Fig.~\ref{fig:soar_flight}.  Due to safety constraints associated with the size of the UAV platform, the maximum number of nodes that could be deployed within the netted enclosure was limited to two. 

Although swarm deployment capacity was restricted, the site offered significant practical advantages. Long-term UAV storage and maintenance facilities were located immediately adjacent to the enclosure, which substantially reduced deployment time and enabled rapid evaluation of both device- and network-level configurations.

\textbf{Open-Space Deployment.}  
The large-scale open-air facility was used for the deployment and operation of a swarm of nine UAV nodes, as illustrated in Fig.~\ref{fig:duette_swarm}. While this site provides substantially greater space for swarm experimentation, it also introduces logistical challenges. Transportation of UAV platform and associated equipment required multiple days of travel. Although UAVs were carefully calibrated prior to transport, travel duration and on-road instability necessitated recalibration at the site, significantly increasing network deployment time.

\begin{figure}[t]
    \centering
    \includegraphics[width=0.89\linewidth]{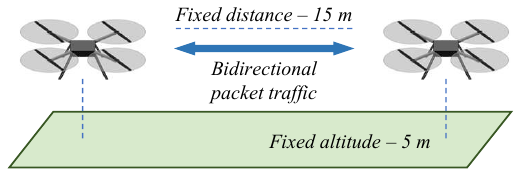}
    \caption{\small Mission 1: Link Optimization.}
    \label{fig:scenario_gain}
\end{figure}

\begin{figure}[t]
    \centering
    \includegraphics[width=0.94\linewidth]{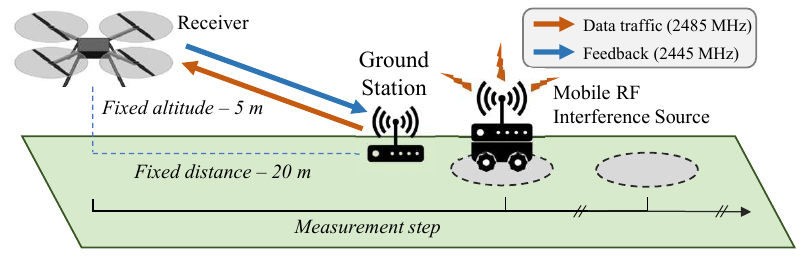}
    \caption{\small Mission 2: Interference Localization. \vspace{-4mm}}
    \label{fig:scenario_jammer}
\end{figure}

\subsection{Experimental Methodology}\label{sec:experimental_methodology}
The UAV swarm network described in Sec.~\ref{sec:swarm_development} was utilized to demonstrate the C2Stack framework prototype. For these experiments, the PHY-layer devices were configured to operate in the 2.4~GHz band using two 20~MHz channels centered at 2.485~GHz and 2.465~GHz, with 64-QAM modulation.

The UAV deployment topology for Mission~1 is illustrated in Fig.~\ref{fig:scenario_gain}. Through experimentation, the functional range of $G_{rx}$ was determined to be 15–70~dB. In certain trials, the action space $\mathcal{A}$ was defined as a subset of this range in order to evaluate the impact of action-space size on the convergence rate of the MIX-MAB algorithm.

For Mission~2, the network deployment scheme is shown in Fig.~\ref{fig:scenario_jammer}. In this setup, an aerial node hovers at a fixed altitude of 5~m and is positioned 20~m from a continuously transmitting ground node. RF interference is generated using the software-defined interference generator, which is placed at varying distances from the aerial receiver. The interference profile was optimized experimentally by comparing network performance under different conditions. Results show that the most disruptive interference was caused by intermittent transmission of a 1~MHz narrowband BPSK waveform. 

During transmission, network performance metrics were collected at the aerial receiver following the process outlined in Sec.~\ref{sec:interference_localization}. We generate 6 such datasets for offline inference, which we refer to as datasets A, B, C, D, J, and K, respectively, across the two test sites. Specifically, datasets A, B, C, and D are collected in the netted enclosure, and datasets J and K were collected in the larger test facility.

\begin{figure}[t]
    \centering
    \includegraphics[width=0.98\linewidth,height=0.55\linewidth]{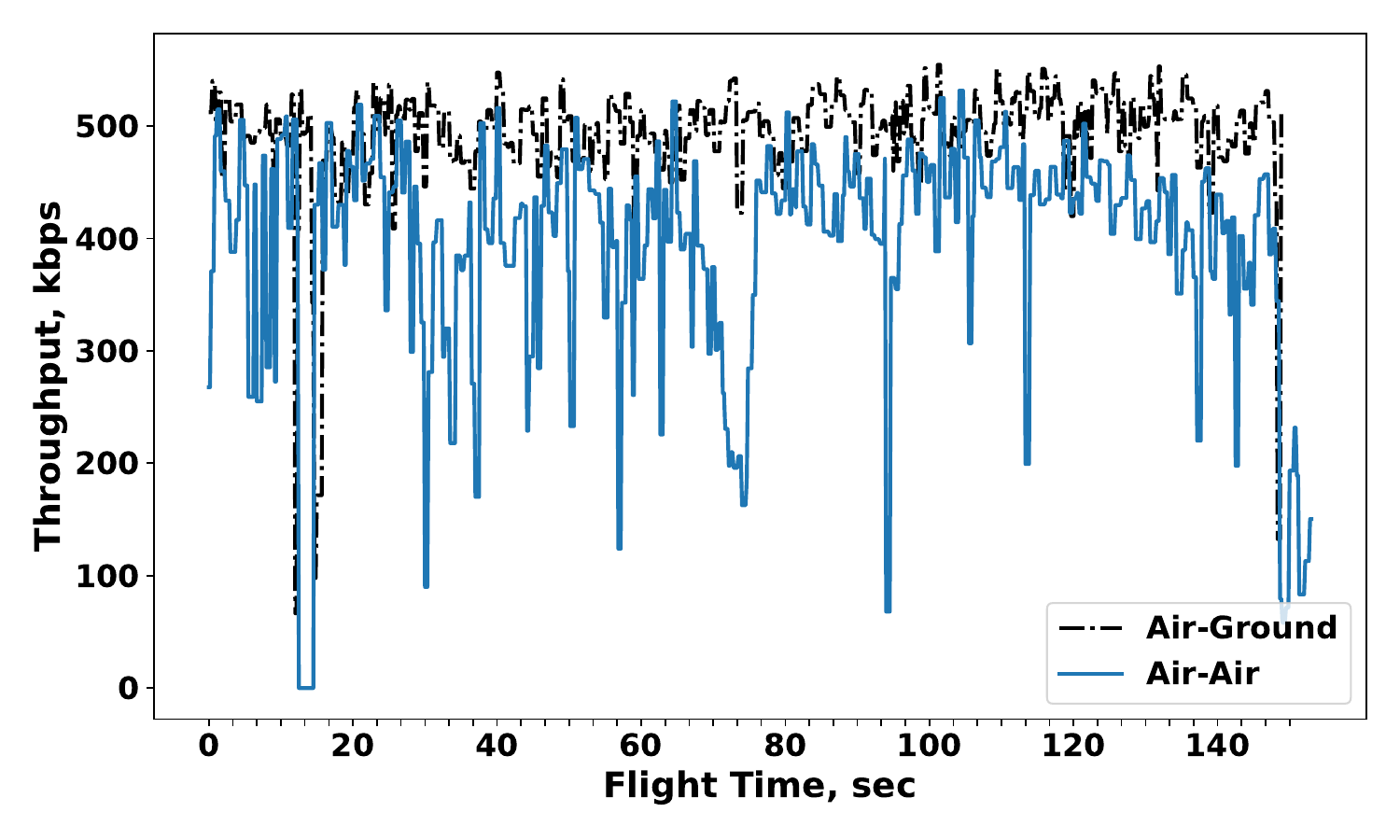}
    \vspace{-2mm}
    \caption{\small Comparison of throughput during flight tests, considering A2A (blue) and A2G (black) links. }
    \vspace{-3mm}
    \label{fig:throughput_compare}
\end{figure}

\subsection{Measurement and Analysis}
\textbf{Initial Measurements.}  
Before executing the two missions described in Sec.~\ref{sec:mission}, a series of flight tests were conducted to validate the C2Stack prototype and to establish baseline heuristics across key configuration variables, including modulation, transmit and receive gain, and range, considering both A2A/A2G links. 
We include an example of these measurements to illustrate one of the key deployment challenges in the establishment of C2Stack. Fig.~\ref{fig:throughput_compare} shows throughput measurements collected at the netted enclosure test site. UAV nodes were set to hover at 12 meters altitude, and in both cases transmitter and receiver were deployed at 30 meters horizontal distance. 
We have also profiled the packet latency across both software and hardware components (see Appendix A.1 for further details) to provide a benchmark for signal propagation delays.
These initial evaluations provided valuable insights regarding link stability, feedback latency, and deployment limitations that informed the experimental procedures presented later in this section and will guide future investigations using the C2Stack framework.

As shown in Fig.~\ref{fig:throughput_compare}, the observed throughput of both A2A and A2G links is highly dynamic during network operation. Although this is not surprising, we emphasize this as a key challenge for optimization tasks that rely on network-level observations. In particular, although many studies model A2A and A2G channels using noise distributions, our results indicate that even under static deployments, network performance may deviate from standard distributions and instead require models with time-varying parameterization. 
As indicated in Appendix A.2, which provides an ideal upper-bound estimate of network scalability, this challenge is further complicated by large-scale swarm development, and advanced spectrum management strategies may be required to maintain link reliability as network size increases. A more comprehensive investigation of network scalability will be conducted in future work using a fleet of 30 UAVs which is currently under development.

\subsection{C2Stack Measurement Results} \label{sec:cstackResult}
In this section, we present measurement results for the missions introduced in Sec.~\ref{sec:mission}, demonstrating how C2Stack supports diverse data-driven modeling and control tasks.

\begin{figure}[t]
\centering
\includegraphics[width=0.99\linewidth,height=0.52\linewidth]{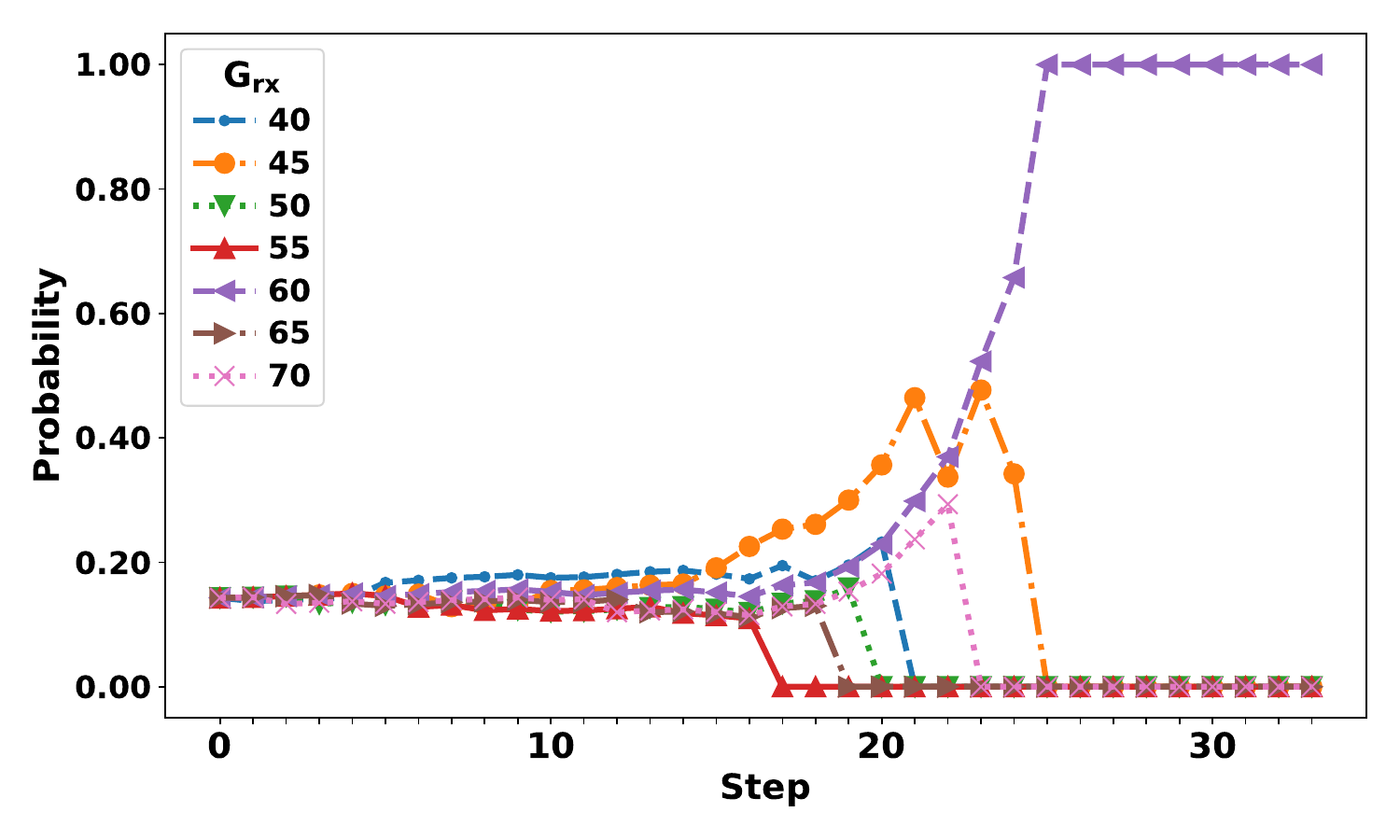} 
\vspace{-3mm}
\caption{\small Action probabilities during protocol optimization, indicating selection of optimal $g_{k}=60$ after $\tau=25$. \vspace{-4mm}
}
\label{fig:opts_1_best}
\end{figure}

\textbf{Mission 1.}  
We begin by examining the ideal online convergence behavior of the MIX-MAB algorithm. 
In order to ensure successful OTA operation, Mission 1 was first tested using a hardware-in-the-loop (HITL) channel emulator (see Appendix B for further details) to validate the algorithm integration logic offline, prior to A2A evaluation. This further allowed us to derive a rough estimate of desirable parameters by excluding any $g_{k}\in G_{rx}$ which resulted in zero or low link throughput. We note that the HITL testbed was used only for debugging and implementation steps, and that all results presented in this section reflect online performance in A2A field trials. As shown in Fig.~\ref{fig:opts_1_best}, the algorithm successfully identifies the optimal value of $g_{k}=60$~dB by monitoring network performance, following the framework described in Sec.~\ref{sec:protocol_optimization}, and removes sub-optimal actions by setting their probability to 0 by $\tau=25$. 

\begin{figure}[t]
  \centering
    \begin{tabular}{cc}
    \hspace{-2mm}\includegraphics[width=0.48\linewidth]{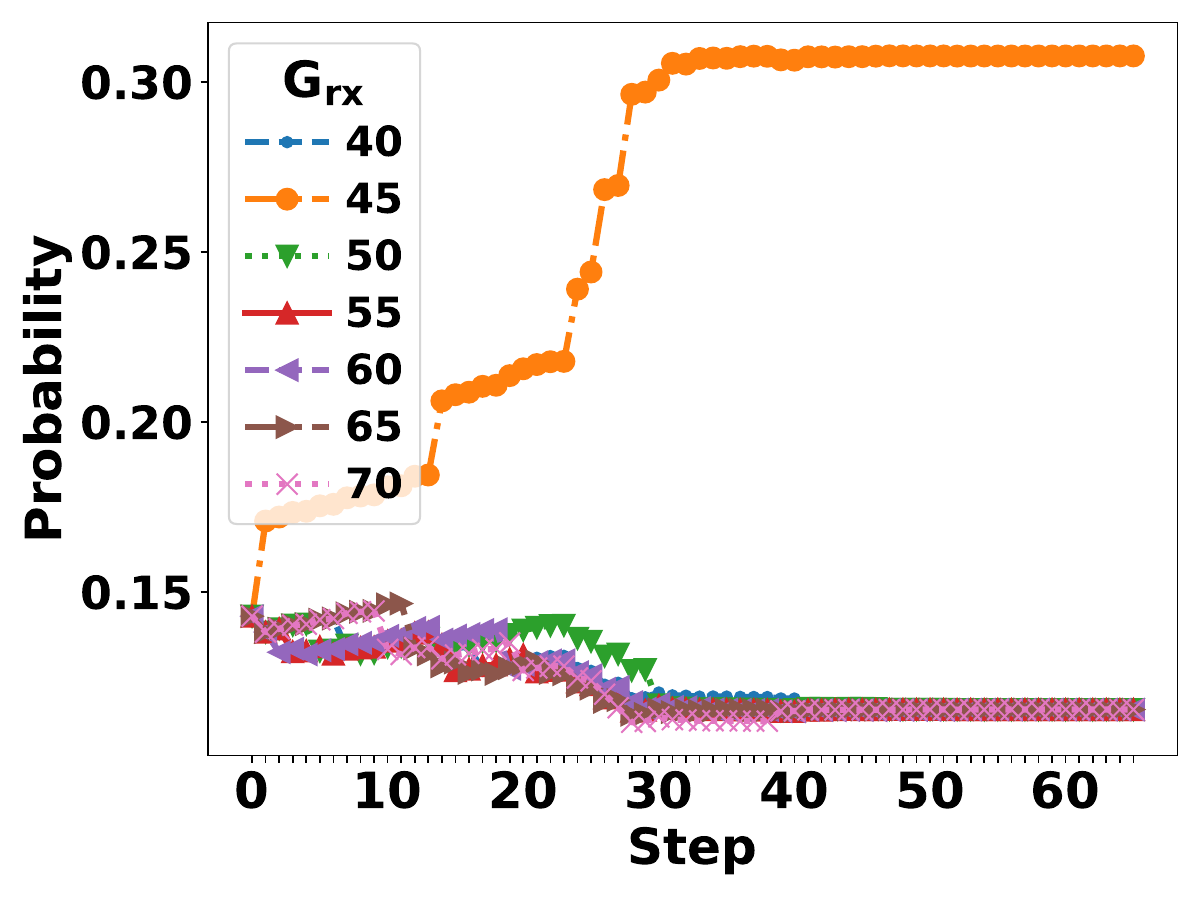} &
    \hspace{-3mm}\includegraphics[width=0.48\linewidth]{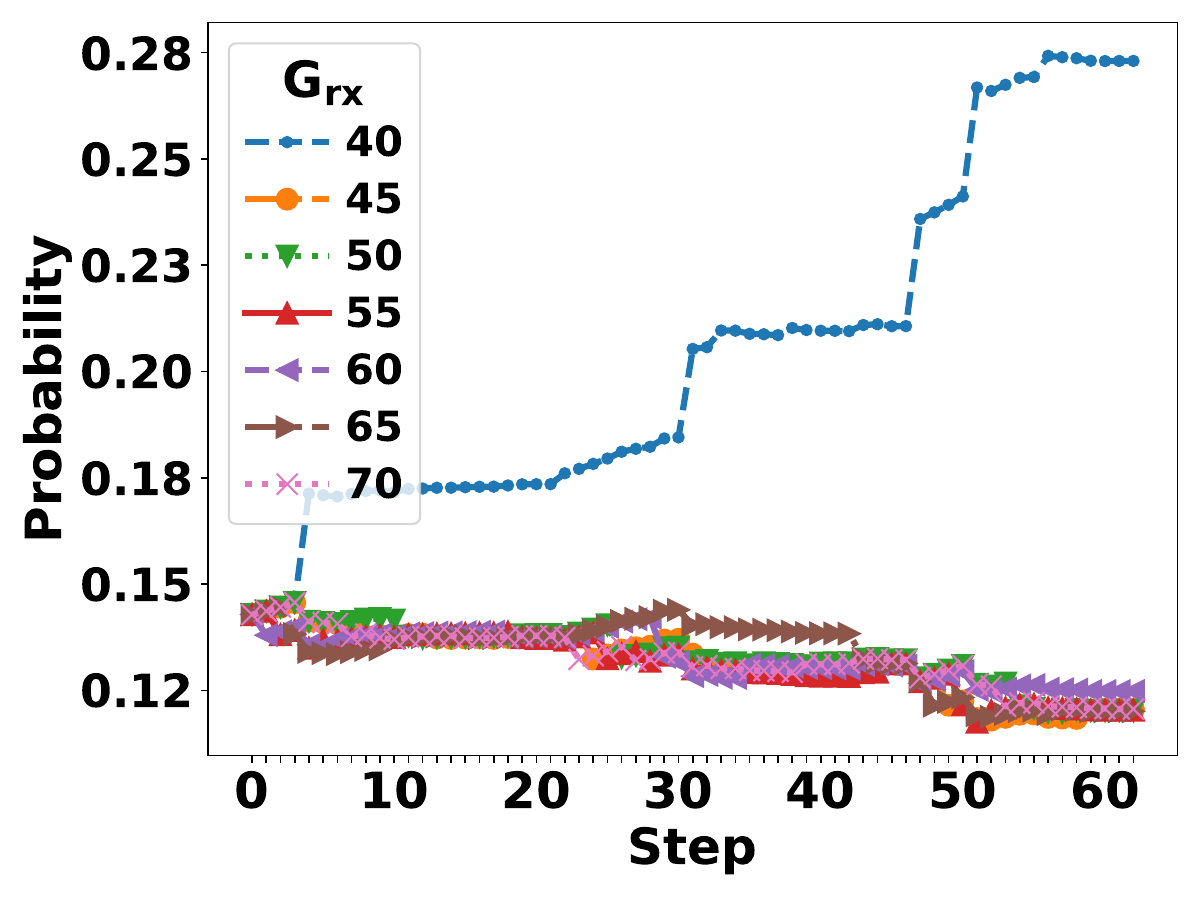} \\[-4pt]
    \small (a)  & \small (b) \\[-4pt]
  \end{tabular}
  \vspace{-1mm}
  \caption{\small Simultaneous execution of protocol optimization algorithm, showing action selection probability during flight test for paired nodes: (a) node 1; and (b) node 2.}
  \label{fig:opt_1_2_compare}
\vspace{-1mm}
\end{figure}

\begin{figure}[t]
  \centering
  \begin{tabular}{cc}
    \hspace{-2mm}\includegraphics[width=0.48\linewidth]{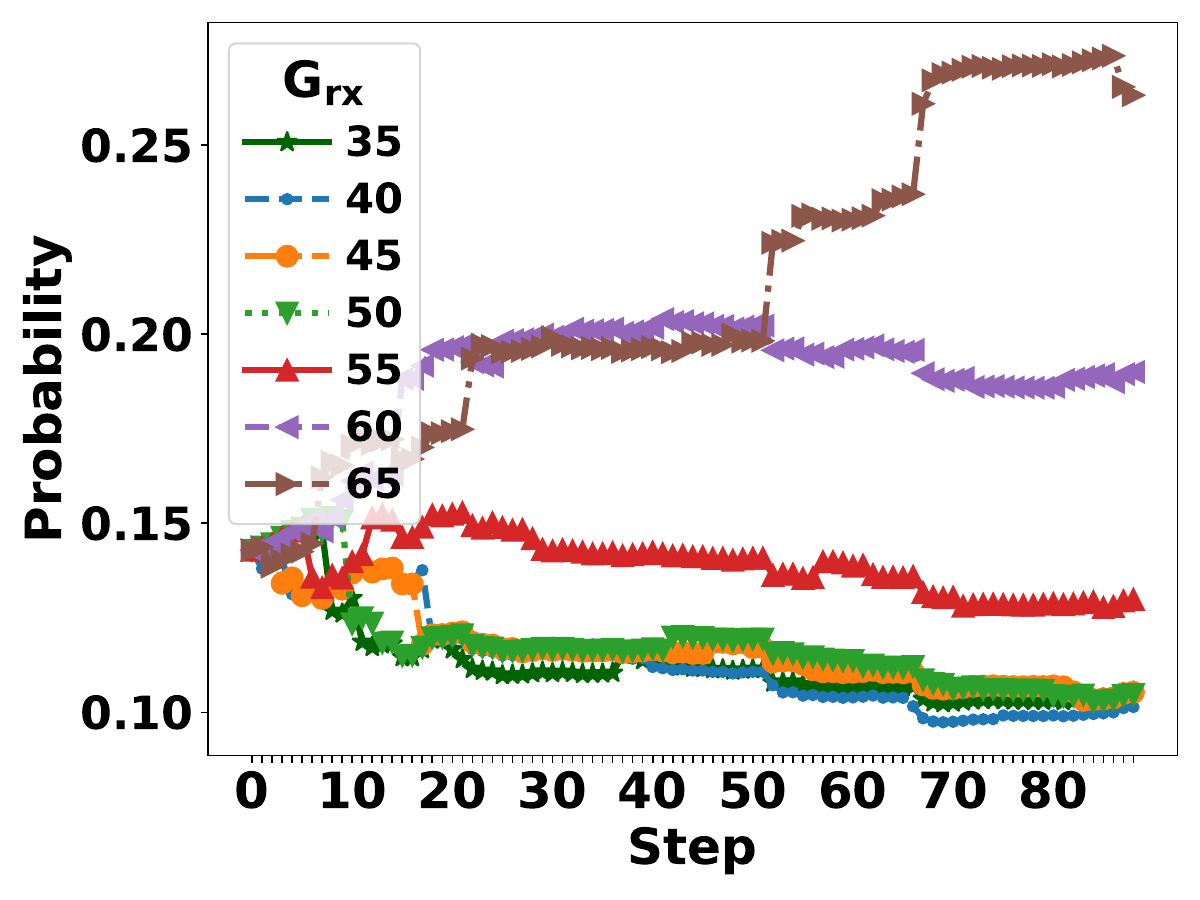} &
    \hspace{-3mm}\includegraphics[width=0.48\linewidth]{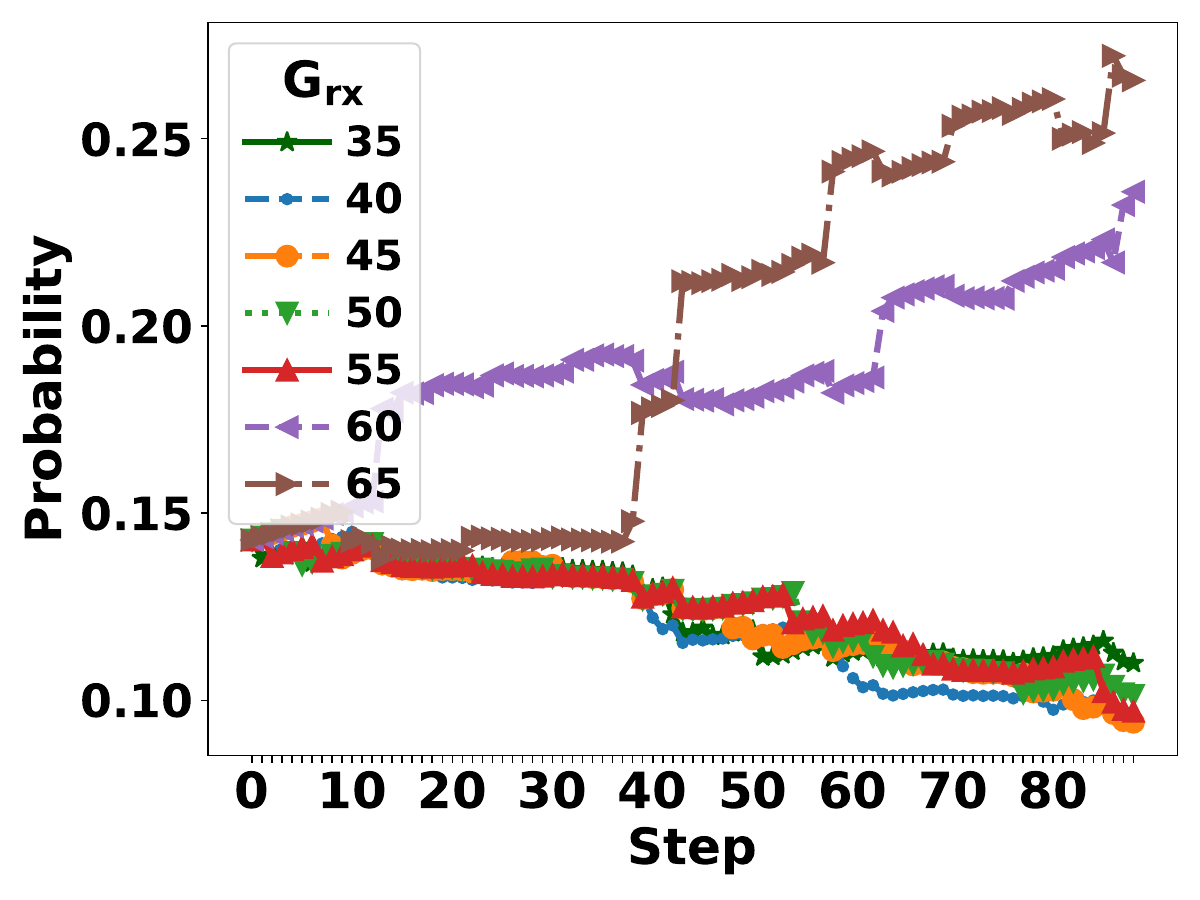} \\[-4pt]
    \small (a)  & \small (b) \\[-4pt]
  \end{tabular}
  \vspace{-1mm}
  \caption{\small Simultaneous execution of protocol optimization algorithm, showing action selection probability during flight test for paired nodes: (a) node 2; and (b) node 3. 
  }
  \label{fig:opt_2_3_compare}
    \vspace{-1mm}
\end{figure}

We further compare online convergence outcomes across multiple UAV nodes. Figs.~\ref{fig:opt_1_2_compare} and \ref{fig:opt_2_3_compare} show optimization at both ends of an A2A link simultaneously.
The agent in Fig.~\ref{fig:opt_1_2_compare}(a) begins converging to $g_{k}=45$~dB, which achieves a $P_{k}^{m}(\tau)$ of 0.31 after $\tau=30$, with all other $g_{k}\in G_{rx}$ reaching a $P_{k}^{m}(\tau)\leq0.11$ after $\tau=41$. Similarly, the agent in Fig.~\ref{fig:opt_1_2_compare}(b) begins converging to $g_{k}=40$~dB, with $P_{k}^{m}(\tau)$ reaching 0.27 by $\tau=56$, although neither agent was able to remove sub-optimal actions within $T_{m}$. The difference in selected $g_{k}$ in this scenario emphasizes the complexity of the A2A channel. 
While the nodes in Fig.~\ref{fig:opt_1_2_compare} start to converge toward single values of $g_{k}$, the nodes in Fig.~\ref{fig:opt_2_3_compare} indicate convergence toward multiple values. Fig.~\ref{fig:opt_2_3_compare}(a) and Fig.~\ref{fig:opt_2_3_compare}(b) indicate favorable selection by both agents of $g_{k}=60$~dB and $g_{k}=65$~dB, however neither achieves $P_{k}^{m}(\tau)$ above 0.28 within $T_{m}$. Both Fig.~\ref{fig:opt_1_2_compare} and Fig.~\ref{fig:opt_2_3_compare} demonstrate that high variance of the A2A channel can make agent rewards inconsistent, resulting in unstable online convergence.

\begin{figure}[t]
  \centering
  \begin{tabular}{cc}
    \hspace{-2mm}\includegraphics[width=0.48\linewidth]{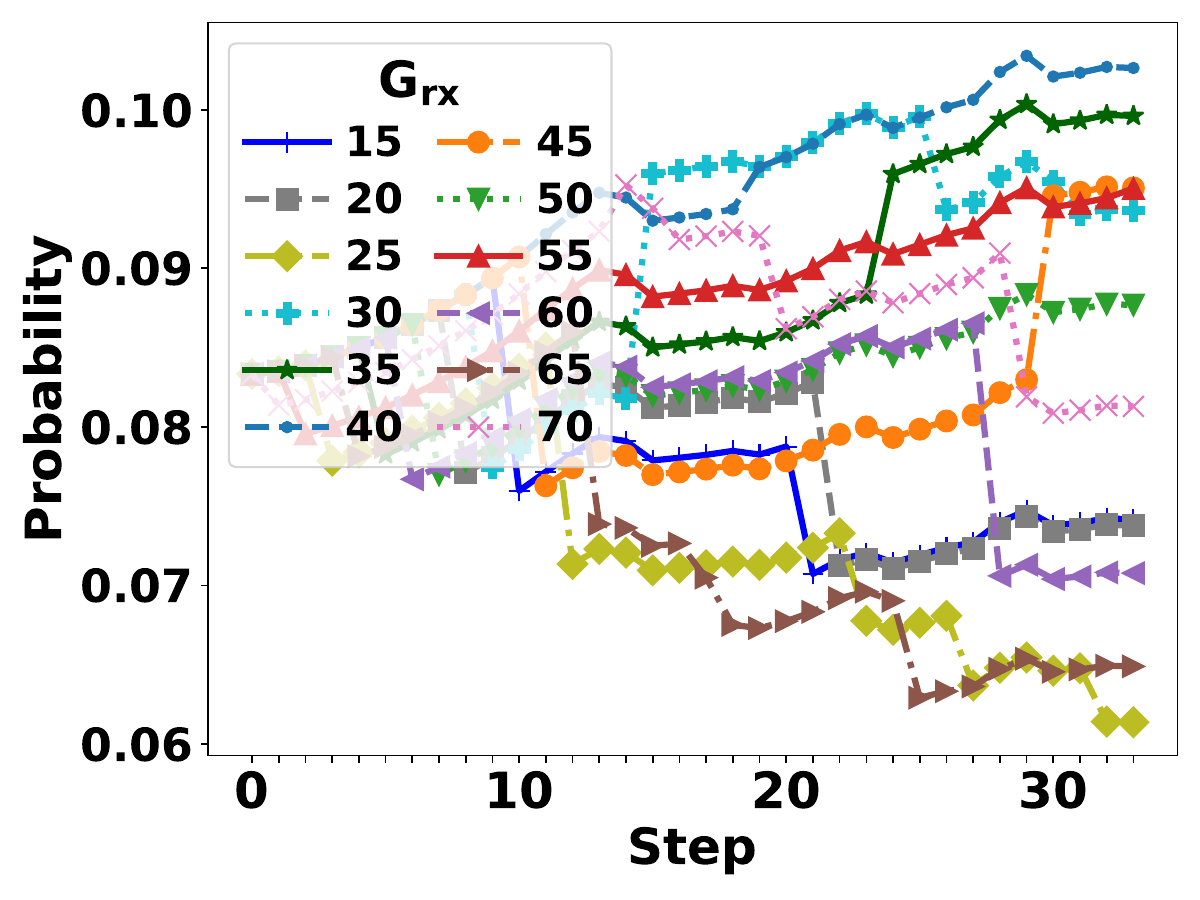} &
    \hspace{-2mm}\includegraphics[width=0.48\linewidth]{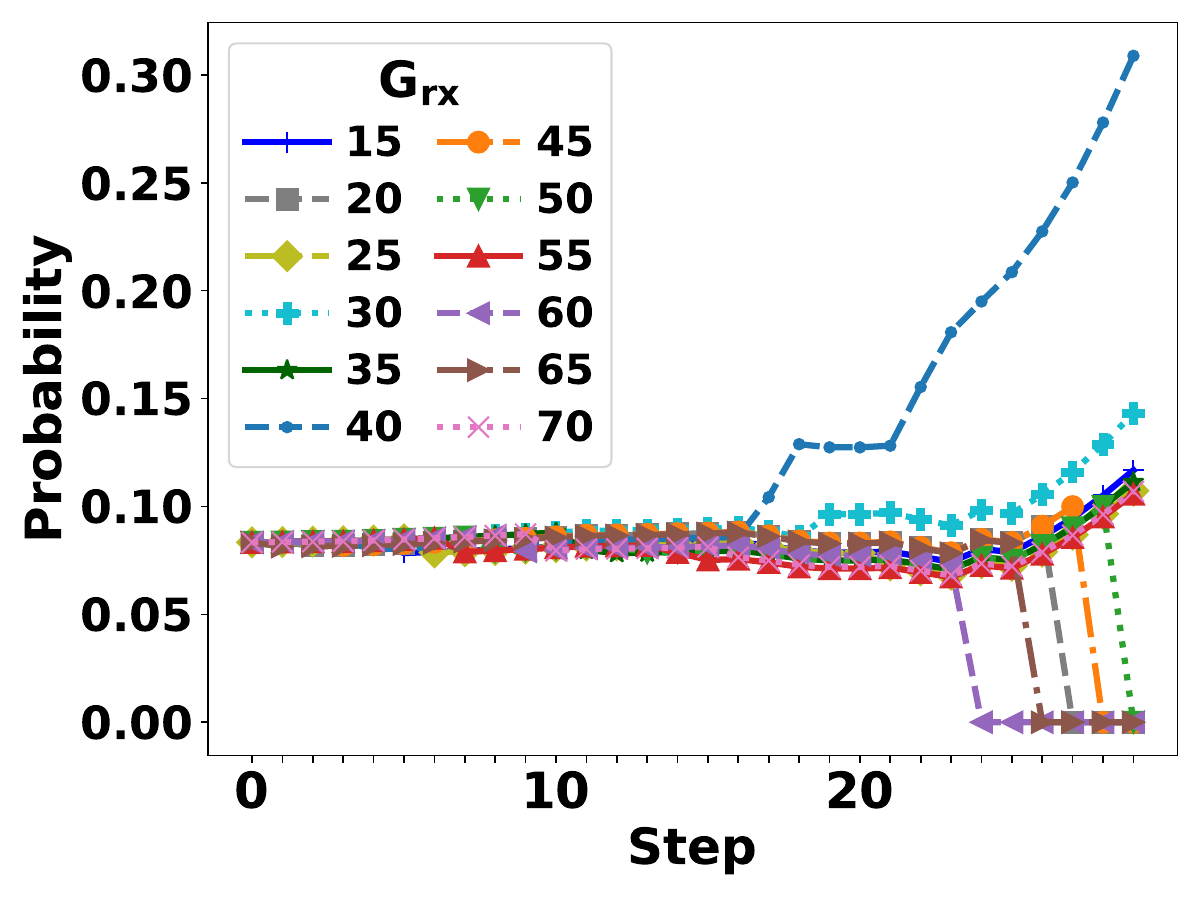} \\[-4pt]
    \small (a)  & \small (b) \\[-4pt]
  \end{tabular}
  \vspace{-1mm}
  \caption{\small Optimization performance of consecutive trials in expanded action space: (a) trial 1; (b) trial 2. 
  }
  \label{fig:opt_2_actionspace}
    \vspace{-1mm}
\end{figure}

We then investigate the impact of enlarging the action space from $K=7$ to $K=12$ on online learning performance. Two representative cases are shown in Fig.~\ref{fig:opt_2_actionspace}: (a) persistent uncertainty, reflecting dynamic channel conditions and high variance in rewards; and (b) delayed convergence, where the optimal value is eventually identified. The learning process shown in Fig.~\ref{fig:opt_2_actionspace}(a) is chaotic, which implies variance among observed $\Bar{R}_{m}(\tau)$ at each $g_{k}$ hence inconsistency among rewards. The agent in Fig.~\ref{fig:opt_2_actionspace}(b) identifies $g_{k}=40$~dB as the optimal case, achieving $P_{k}^{m}(\tau)=0.30$ by $T_{m}$ while starting to remove sub-optimal values at $\tau=24$. The consecutive trials in Fig.~\ref{fig:opt_2_actionspace}(a) and Fig.~\ref{fig:opt_2_actionspace}(b) are conducted less than 15 minutes apart, emphasizing temporal inconsistency in the observed performance of A2A link configurations.

\begin{figure}[t]
  \centering
  \begin{tabular}{ccc}
  \multicolumn{3}{c}{\includegraphics[width=0.96\linewidth]{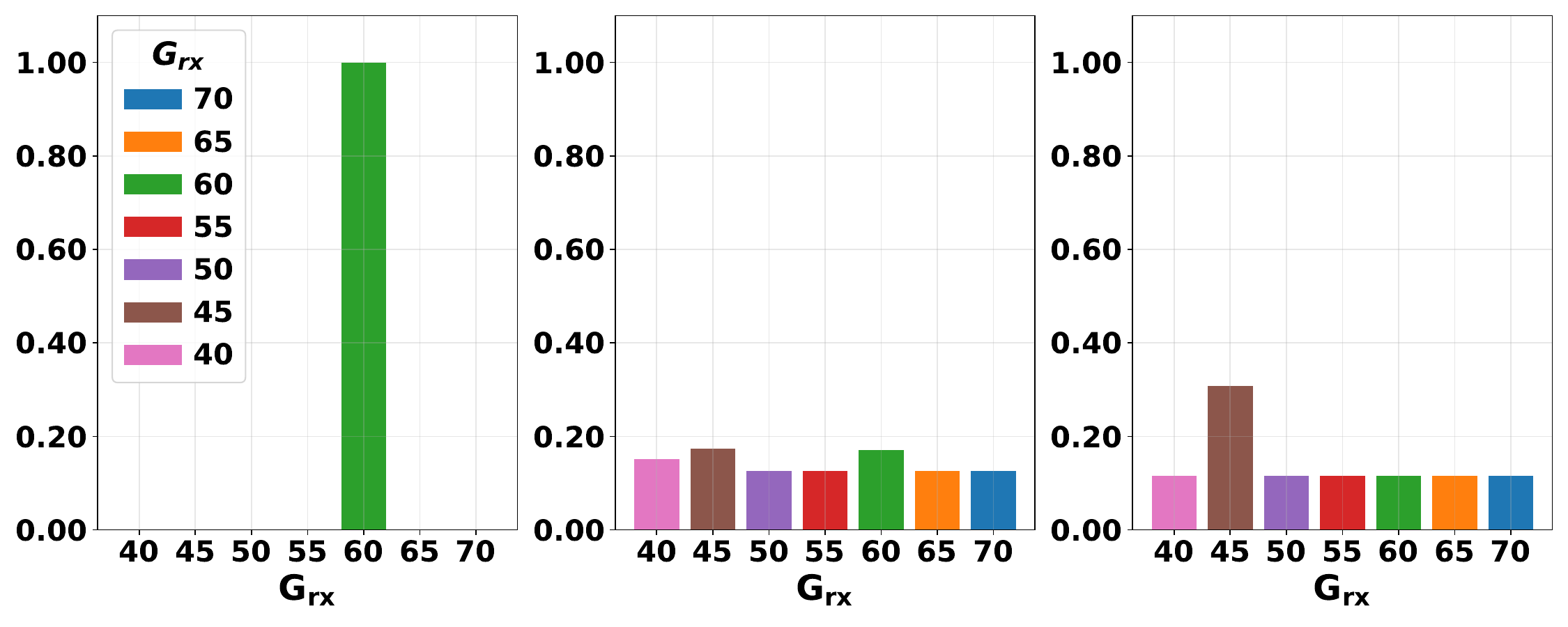}} \\[-4pt]
  \makebox[0.36\linewidth]{\small (a)} & 
  \makebox[0.2\linewidth]{\small (b)} & 
  \makebox[0.2\linewidth]{\small (c)} \\[-4pt]
  \end{tabular}
  \vspace{-1mm}
  \caption{\small Final action probability distributions of trials in consecutive flight tests: (a) trial 1; (b) trial 2; and (c) trial 3. \vspace{-3mm}}
  \label{fig:node_1_combined}
\vspace{-1mm}
\end{figure}

Finally, Fig.~\ref{fig:node_1_combined} presents final distribution of $P_{k}^{m}(\tau)$ following online iteration for consecutive trials. In Fig.~\ref{fig:node_1_combined}(a), the agent converges completely to $g_{k}=60$~dB, removing all other actions. By contrast, Fig.~\ref{fig:node_1_combined}(b) shows a nearly uniform distribution across actions, while Fig.~\ref{fig:node_1_combined}(c) suggests convergence toward $g_{k}=45$~dB. These results further demonstrate the temporal inconsistency of A2A channel effects, implying A2A links must be continually optimized to adapt to time-varying channel characteristics.

Overall, these results demonstrate that while MIX-MAB can successfully identify optimal parameter values, convergence behavior depends strongly on action space size and time-varying network conditions during operation. In general, due to the highly dynamic nature of A2A channels, including rapidly changing link quality, intermittent connectivity, and non-stationary interference, direct deployment of conventional RL algorithms can lead to unstable or inconsistent convergence. To address this challenge, the CNOS module of C2Stack monitors key mission metrics for diagnostic logging, including packet delivery ratio (PDR), link quality indicators, and reward trajectories. This can enable evaluation of adaptive training strategies, such as policy rollback, divergence detection, or parameter adaptation, as well as provide insight on the limitations of online RL in UAV networks. A detailed investigation of these approaches is beyond the scope of this work and left for future study.

\begin{figure}[t]
  \centering
  \begin{tabular}{ccc}
    \hspace{-3mm}\includegraphics[width=0.33\linewidth]{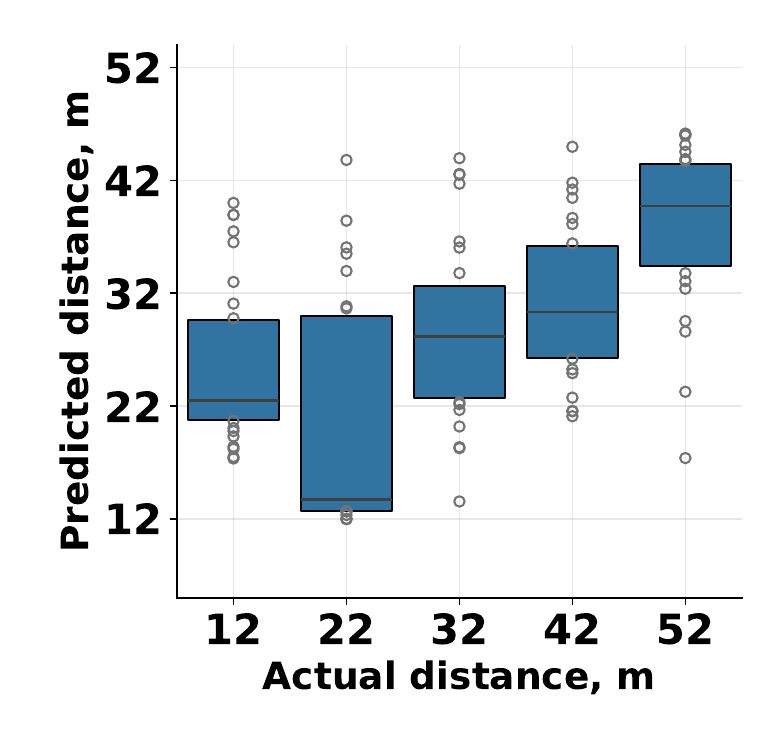} &
    \hspace{-4mm}\includegraphics[width=0.33\linewidth]{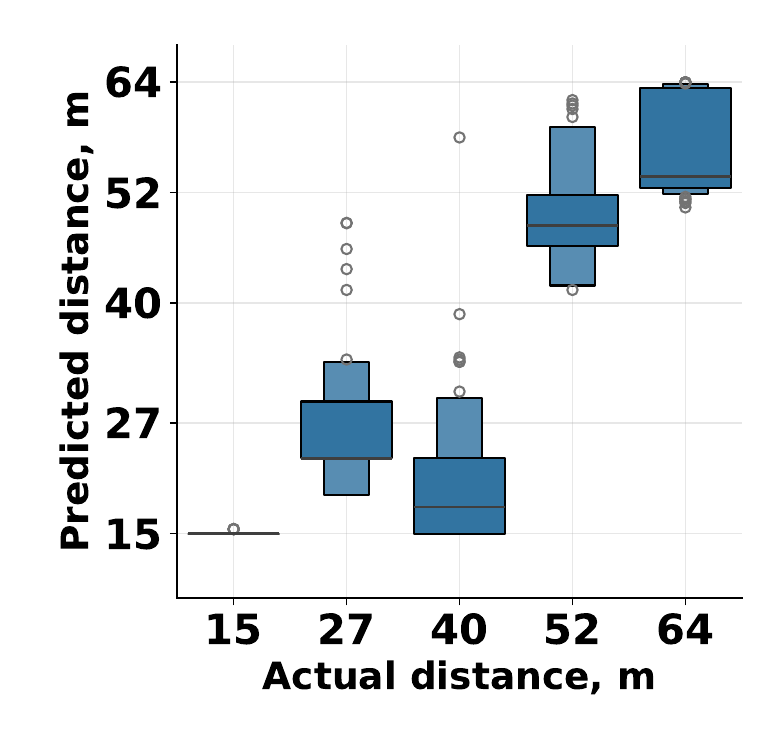} &
    \hspace{-4mm}\includegraphics[width=0.33\linewidth]{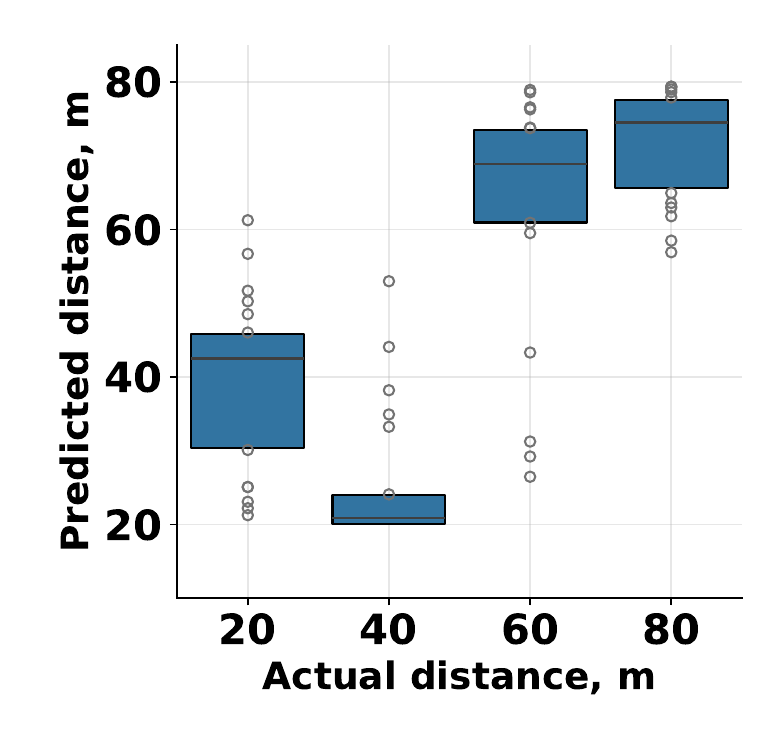}\\[-4pt]
    \small (a)  & \small (b) & \small (c) \\
  \end{tabular}
  \vspace{-1mm}
  \caption{\small Interference source distance estimation accuracy, based on various datasets: (a) training set A testing set B; (b) training set C testing set D; (c) training set J testing set K.}
  \label{fig:distance_est}
  \vspace{-1mm}
\end{figure}

\begin{figure}[t]
  \centering
  \begin{tabular}{cc}
    \hspace{-2mm}\includegraphics[width=0.48\linewidth]{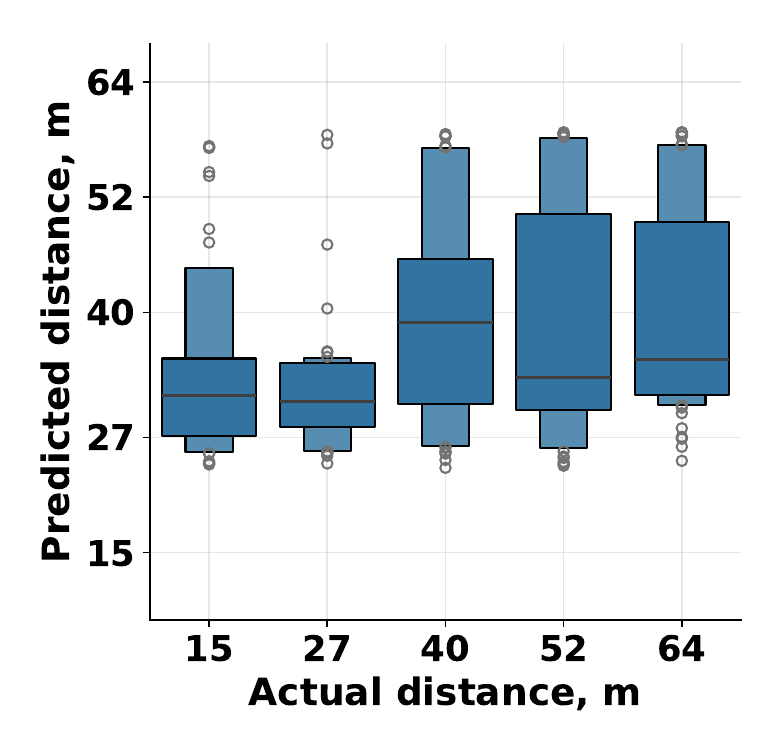} &
    \hspace{-2mm}\includegraphics[width=0.48\linewidth]{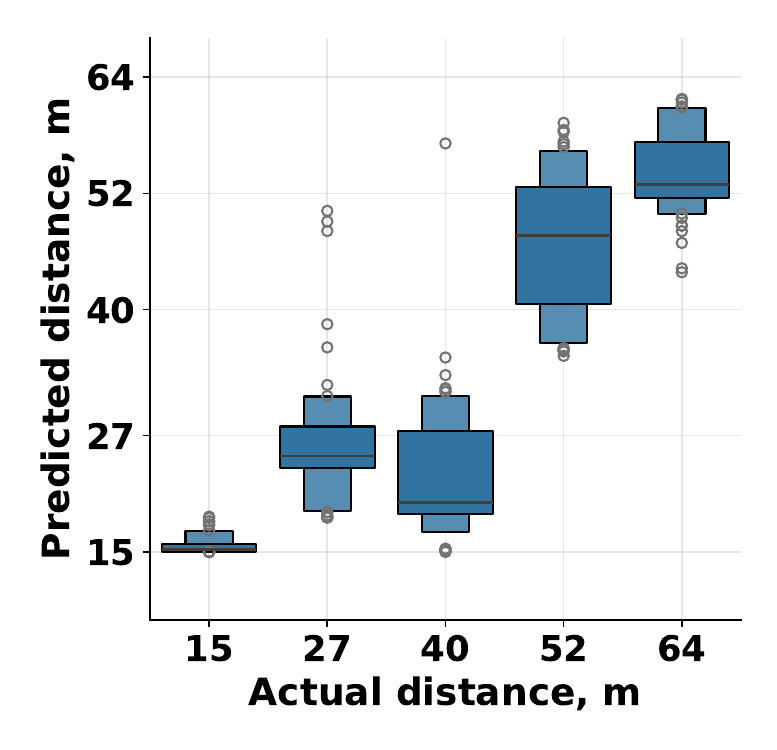} \\[-4pt]
    \small (a)  & \small (b) \\[-4pt]
  \end{tabular}
  \caption{\small Offline interference estimation accuracy considering offline bootstrap: (a) training set A, B, C testing set D; (b) training set J, K, C testing set D. \vspace{-3mm}}
  \label{fig:distance_inference}
\end{figure}

\textbf{Mission 2.}
Fig.~\ref{fig:distance_est} illustrates the performance of the random forest regression model used for offline inference of $d_{m}(\tau)$, using the datasets described in Sec.~5.2. We first consider inference results from the initial datasets A and B, shown in Fig.~\ref{fig:distance_est}(a), where the model demonstrates accurate estimation across most $d_{m}(\tau)$. Inference across the second datasets, C and D, shown in Fig.~\ref{fig:distance_est}(b), also achieve reliable predictions, though reduced accuracy is observed at $d_{m}(\tau)=40$~m. Finally, the inference across large-scale datasets J and K collected in the larger test facility, presented in Fig.~\ref{fig:distance_est}(c), exhibits similar behavior: prediction uncertainty increases at 40~m, while the most accurate estimations occur at 80~m. Overall, these results highlight the effectiveness of the proposed approach in estimating $d_{m}(\tau)$ without dedicated spectrum sensing hardware, while also indicating that certain ranges of $d_{m}(\tau)$ (e.g., 40~m) pose greater challenges and require further investigation.

Fig.~\ref{fig:distance_inference} presents the offline inference performance of the IDLS using historical datasets during training. Since the IDLS operates as a C2-App, its offline inference performance is identical to online operation, as the offline datasets were streamed into the IDLS in the same format as real-time network data. Fig.~\ref{fig:distance_inference}(a) shows the inference of observations in dataset B of a model trained on datasets J, K, and A. The inclusion of larger $d_{m}(\tau)$ values reduces inference accuracy overall. By contrast, Fig.~\ref{fig:distance_inference}(b) shows the inference of observations in dataset D of a model trained on datasets J, K, and C. In this case, the additional training data improves prediction accuracy at the $d_{m}(\tau)=40$~m, though it introduces slightly higher variance across other ranges.

\vspace{-3mm}
\section{Lessons Learned and Challenges}\label{sec:lessons}

While the C2Stack framework has proven to be a useful tool for rapid deployment and evaluation of data-driven control algorithms in swarm networking, our experience with the prototype introduced in Sec.~\ref{sec:framework} has revealed several experimental limitations that point to important directions for future improvement.  

\textbf{Deployment Area Requirements.}  
The UAV platforms described in Sec.~\ref{sec:swarm_development} require relatively large open spaces, which complicates the scheduling and design of flight tests. Many urban and semi-urban environments do not provide sufficient mobility to evaluate swarm navigational policies, and FAA regulations further constrain the feasible deployment areas.  

\textbf{Sensing Capability.}  
While the PHY-layer protocol presented in Sec.~\ref{sec:c2stack_pps} provides a rich feature set for UAV communications research, the current design lacks in-situ RF sensing. Without real-time environmental monitoring, C2Stack cannot generate reliable spectral profiles of the deployment environment or adapt protocol configurations accordingly.  

\textbf{Flight Duration.}  
Reliance on battery power strictly limits UAV operation, requiring frequent interruptions to land devices and replace batteries. In our experiments, flight times were restricted to approximately 15 minutes due to battery constraints and the weight of the C2Stack payload. This disruption complicates the training and evaluation of data-driven algorithms. While tethering UAVs to ground-based power stations is possible, this reduces deployment flexibility and adds payload weight, altering flight performance.  

\textbf{Weather Conditions.}  
Weather presents another limitation for UAV networking research. While commercial waterproof UAVs exist \cite{fairman2024waterproofuav}, they currently lack the payload capacity and features required for our experimental platform. As a result, experiments must often be scheduled around weather windows, reducing opportunities for extended flight testing.  

\textbf{Software-Defined Protocol Stack.}  
The PPS prototype, described in Sec.~\ref{sec:pps_config}, is implemented primarily in Python, with some performance-critical modules in C. While this approach accelerates development and supports modularity, it can introduce computational delays compared to lower-level implementations or direct hardware realization. This may pose limitations for experiments requiring high throughput, low latency, or strict timing guarantees \cite{nanz2015languages}. 
Additionally, it is worth clarifying that network security protocols are not included in the base PPS configuration, and deployment of user-defined algorithms may result in vulnerability to various forms of network disruption not investigated in this work, including spoofing, data injection, and targeted denial-of-service (DoS). Once open-sourced, we plan to work with the community to identify possible vulnerabilities and develop countermeasures to ensure the security and integrity of the C2Stack framework.

\textit{Summary:} These limitations provide a roadmap for advancing the C2Stack framework. Future versions will incorporate RF sensing to improve spectral awareness, hardware redesigns to extend UAV flight time, and extensive benchmarking to identify and address bottlenecks in the PPS implementation for low-latency operation. We also envision closer collaboration with the broader research community to continue scaling aerial networking experiments and refining the C2Stack platform for real-world deployment.

\subsection{Research Challenges}\label{sec:challenges}
As discussed in Sec.~\ref{sec:intro}, the evaluation of data-driven modeling and control techniques in real-world scenarios is paramount to advancing the technical readiness of UAV-enabled and UAV-assisted network architectures. While such evaluation in wireless domains is already challenging, aerial deployment introduces a distinct set of difficulties. Motivated by the limitations presented above, we highlight several broader research challenges to serve as a roadmap for experimental design and UAV-enabled wireless networking.  

\textbf{Algorithm Design.} The computational complexity of data-driven algorithms often requires extensive flight times for training and evaluation. While some prior works account for UAV power consumption and battery limitations \cite{park2023MARLcooperativeUAM}, they frequently overlook environmental factors such as wind, temperature, and spectrum occupancy, which cause discrepancies between simulation results and real-world performance. Consequently, algorithm design must explicitly consider convergence time to ensure feasibility. Because UAV channels exhibit dynamic behavior in both A2A and A2G links, online approaches may benefit from historical datasets or synthetic data generation to accelerate convergence. Offline-trained models typically require significant online tuning to mitigate channel variability and hardware inconsistencies. 
In order to isolate algorithm divergence from adverse environmental effects, additional signal processing strategies can be deployed using I/Q samples forwarded by the SDPI to detect unexpected changes in local RF conditions using data-driven modeling and prediction, statistical channel profiling, or real-time predictive simulation.
Although no universal solution exists, resilient learning paradigms such as meta-learning and domain adaptation offer promising directions to improve adaptability and robustness \cite{mcmanus2022sourcetotarget}.  

\textbf{Scenario Design.} Deploying a UAV swarm network at scale to enable rigorous, reproducible experiments across diverse networking capabilities remains a formidable challenge shared by many research groups. As detailed in Sec.~\ref{sec:swarm_development}, such deployments demand extensive service contracts and legal agreements with research facilities, private landowners, and public services. For academic groups bound by university policies, this creates substantial barriers in time, cost, and administration. 
Relocating hardware to diversify experimental contexts amplifies these hurdles, requiring renewed agreements as well as complex logistics for hardware and personnel transport. Additionally, weather uncertainty further complicates the orchestration of large-scale UAV experiments, underscoring the inherent fragility of such research infrastructures. The AERPAW platform \cite{marojevic2020aerpaw} seeks to alleviate some of these challenges, providing the facilities, hardware, and expertise for UAV experiments.
However, the platform does not support scenarios that can be configured flexibly for dynamic UAV path planning or network control decisions, restricting the scope of enabled research. 

\textbf{Hardware Design.} The design and integration of UAV hardware platforms critically influence the capabilities of UAV-enabled networking systems. UAV airframes introduce non-trivial blockage effects \cite{ali2024losprobability}. The vibration and instability of UAVs degrade network performance and pose risks to operational safety \cite{qi2024vibrationmodel}, \cite{dou2024wobble}. Effective hardware design must prioritize stability and resilience while supporting diverse networking functionalities. Payloads should be designed to provide rich diagnostic data and system-wide monitoring, enabling rapid fault detection and behavioral analysis. Such features are essential to ensure that UAV platforms not only sustain network performance but also serve as robust experimental testbeds for advancing UAV networking research.  

\section{Conclusions}\label{sec:conclusion}
We have introduced C2Stack, a swarm control framework that can be used to accelerate the implementation and evaluation of data-driven control algorithms for connected autonomous aerial networks. Its effectiveness has been demonstrated through two benchmark experiments, protocol self-configuration and interference source localization, which highlight the feasibility and value of real-world aerial network experimentation. These results reinforce the need for practical validation of data-driven networking approaches and provide a foundation for future work in this domain. Building on the lessons gained, we have outlined key research challenges in designing more capable, accessible, and flexible UAV-enabled network experimentation platforms.

\bibliographystyle{ieeetr}
\bibliography{c2stack}


\appendix

\section{Appendix: C2Stack System Profiling}\label{app:profile}

We measure and analyze the latency and scalability performance of the integrated C2Stack system to provide reference benchmarks for future experiments conducted using the C2Stack framework. The experiments are conducted in the 2.4~GHz frequency band using a DSSS-based PHY configuration with a chirp rate of 6,666.66~kHz, an up-sampling factor of 6, and a sampling rate of 40~MHz. We consider two key performance metrics for system-level profiling: packet-level latency and network scalability. Packet-level latency is defined as the time required to forward a packet from the transmitter application layer to the receiver application layer. Network scalability is evaluated in terms of the achievable control-signal packet delivery ratio (PDR) as the network size increases.

\subsection{System Latency}

The latency of the C2Stack framework has been characterized based on three processes measured along both the transmit and receive paths: PPS latency, SDPI forwarding latency, and FPGA processing latency. PPS latency is defined as the aggregate latency of all processes across each layer of the protocol stack, measured as the time required to transmit a packet from the application (APP) layer packet generator to the SDPI forwarding interface. We measure PPS latency for endpoint nodes, at which frames are processed by the entire protocol stack, as well as for relay nodes in multi-hop links, at which frames are only handled by PHY, MAC, and NET layer processes. SDPI forwarding latency refers to the time required to transport the packet between the Intel NUC and the FPGA via the Direct Memory Access (DMA) controller on the MPSoC device. FPGA processing latency is defined as the time required to read a frame from the DMA buffer, generate the DSSS waveform, and transmit the resulting signal.

Other potential sources of latency, such as the Ethernet connection between the NUC and MPSoC devices, the 
FPGA Mezzanine Card High Pin Count (FMC-HPC) interface between the MPSoC and RF front-end board, and over-the-air signal propagation, are estimated to be in the nanosecond range and are therefore considered negligible by comparison. The average packet-level latency of each component process for both transmit and receive operations is reported in Table~1. By summing the latency induced by PPS, SDPI, and FPGA processes along both the transmit and receive paths, the results show that the average end-to-end frame latency is approximately 8.12~ms for single-hop links. In multi-hop links, transmit and receive processes at relay nodes will increase this value by roughly 6.21~ms per hop. This level of latency is suitable for most payload applications, such as image or video streaming. For non-payload applications with stricter latency requirements, such as real-time control or coordination tasks, the FPGA transmit and receive latency can be reduced by minimizing the size of the PHY dataframe to the minimum required by the application, at the cost of reducing the theoretical maximum throughput. Similarly, PPS latency can be reduced by decreasing the size of configurable subframes within the PHY dataframe. Finally, the latency of the SDPI forwarding interface is determined by the MPSoC hardware and cannot be further configured.

\begin{table}[H]
\centering
\caption{\small C2Stack Latency Measurements}
\label{tab:latency}
\begin{tabular}{|c|c|c|}
\hline
\textbf{Module} & \textbf{Transmit Path} & \textbf{Receive Path} \\
\hline
PPS (endpoint) & 1.629 ms & 1.419 ms \\
PPS (relay) & 1.073 ms & 0.633 ms \\ 
SDPI & 0.0178 ms & 0.0178 ms \\
FPGA & 2.450 ms & 2.585 ms \\
\hline
\end{tabular}
\end{table}

\vspace{-4mm}

\begin{figure}[ht]
  \centering
  \vspace{-4mm}
  \includegraphics[width=0.98\linewidth]{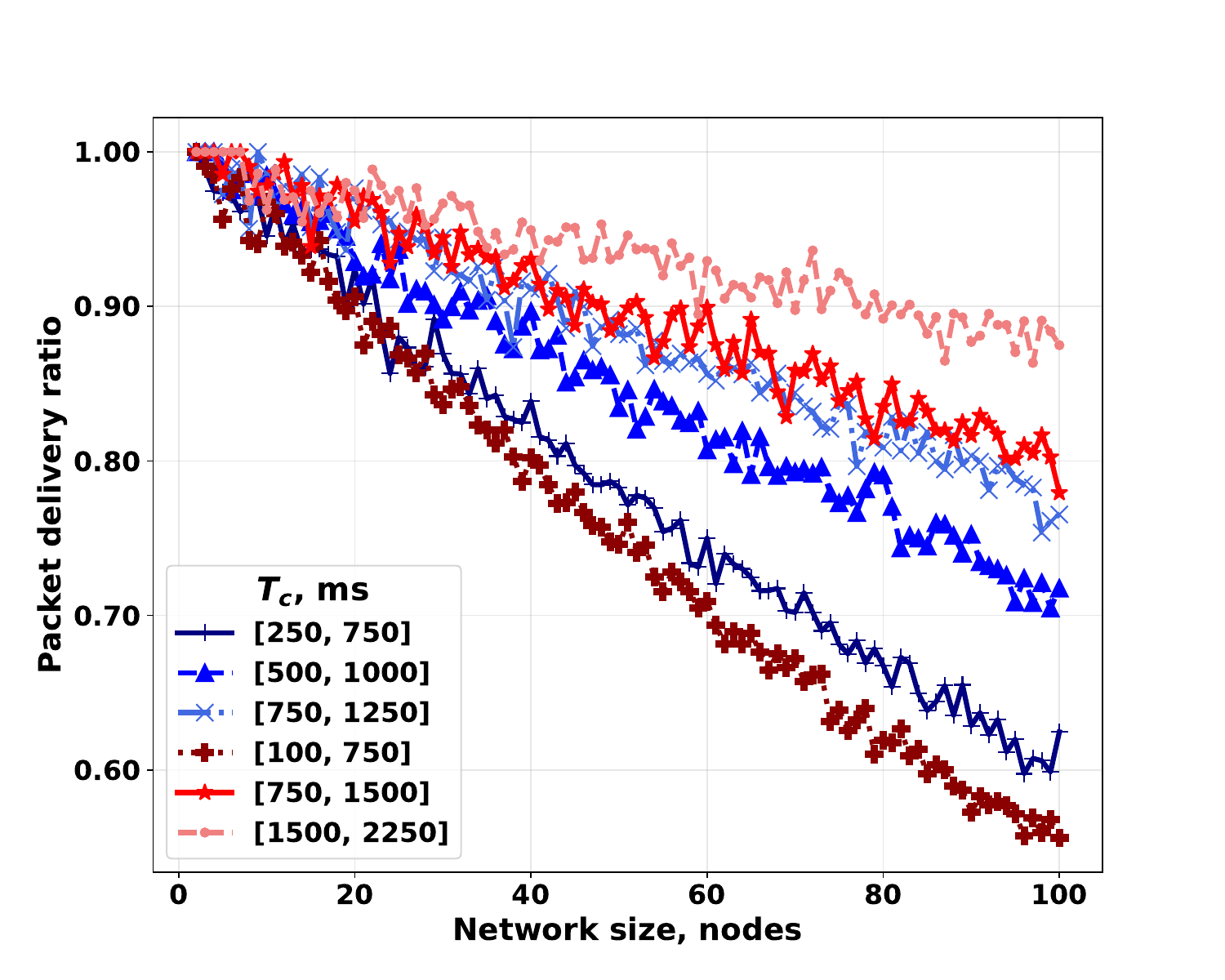}
  \caption{\small Achievable packet delivery ratio (PDR) as a function of network size under different control signal generation intervals.}
  \label{fig:scalability_results}
\end{figure}

\begin{figure}[b]
    \centering
    \vspace{2mm}
    \includegraphics[width=0.98\linewidth]{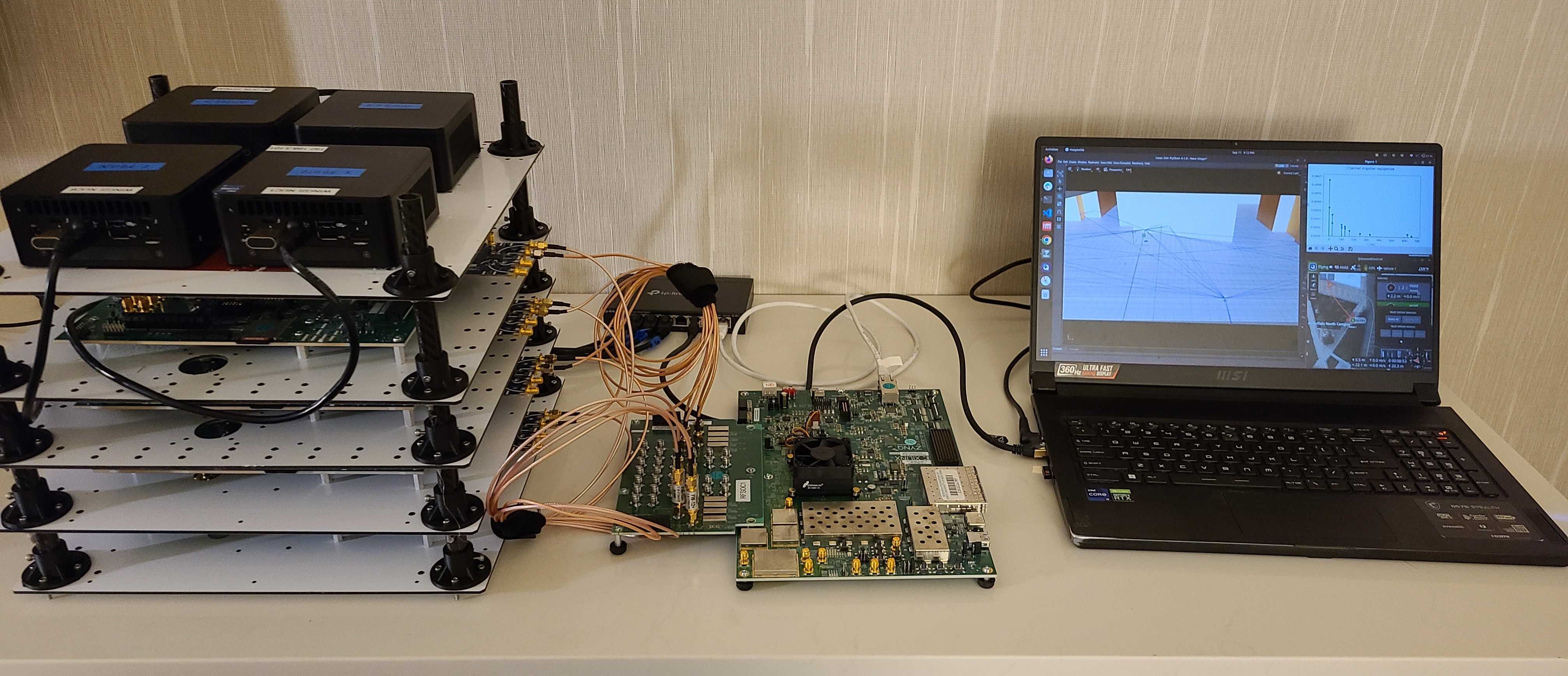}
    \caption{\small Snapshot of HITL Emulator.}
    \label{fig:hitl}
\end{figure}

\subsection{Network Scalability}

To analyze the scalability of the C2Stack framework, we first derive an analytical model for control-plane signaling among nodes in the network. Based on this model, we then develop a lightweight discrete network traffic simulator in Python to estimate channel saturation as the network size $N$ increases. This simulator will be shared under the same open-source license as C2Stack upon release.
Specifically, we initialize an ad hoc network of $N$ nodes, in which each node uses frequency-division duplexing (FDD) to define transmit and receive channels. We define the set of available frequencies as $\mathcal{F} \triangleq \{f_{0}, f_{1}\}$, and alternate the allocation of $f_{0}$ and $f_{1}$ as transmit and receive frequencies according to the index of each node $n \in N$. While a basic channel allocation scheme is considered here, C2Stack is designed to be compatible with more advanced channel allocation strategies that may be developed by researchers.

Control packets are transmitted immediately after generation by their respective generator $G_{c}$, where $c$ denotes the signal context. To mitigate congestion, each $G_{c}$ waits for a random period $t_{c}$ between control signals, defined according to the interval $T_{c}=[t_{c}^{\min}, t_{c}^{\max}]$, in which $t_{c}^{\min}, t_{c}^{\max}$ denote the lower and upper bound values of $t_{c}$, respectively, and must be specified by the user according to service requirements and protocol configuration. The value of $t_{c}$ between generation instances is determined using $t_{c} \sim \mathcal{U}[t_{c}^{\min}, t_{c}^{\max}]$, in which $\mathcal{U}$ denotes the uniform distribution. We measure the packet delivery ratio (PDR), defined as the ratio of successfully received frames to transmitted frames, for network size $N=\{2,\ldots,100\}$. To estimate the upper bound of PDR, we assume a lossless channel and consider frames to be dropped only upon collision. Channel saturation is simulated for a range of $t_{c}^{\min}, t_{c}^{\max}$ values. In the current C2Stack implementation, the interval $T_{Disc}=[500,1000]$ ms corresponds to network discovery signaling during initialization, while the interval $T_{State}=[750,1500]$ ms corresponds to node-state signaling during normal network operation. We consider several other values of $T_{c}$ to represent small variations around these operational settings.

The resulting PDR performance is reported in Fig.~19. As expected, larger and less frequent signaling intervals generally achieve higher PDR as the network size increases, highlighting the tradeoff between network size and control-signal reliability. The lowest PDR occurs when the $T_{c}$ lower bound is less than the size of $T_{c}$, i.e. $t_{c}^{\min} < t_{c}^{\max} - t_{c}^{\min}$, since this configuration results in a higher probability of overlapping transmissions. In the initial C2Stack swarm deployment of 9 nodes, the noted values of $T_{Disc}$ and $T_{State}$ both achieve PDR > 0.95. Since network size and PDR tolerance will vary among use cases, we refer to the provided simulator to estimate optimal specification of $T_{c}$ in future work.

\balance


\section{Appendix: HITL Emulator}

We have developed a hardware-in-the-loop (HITL) channel emulation testbed to accelerate C2Stack framework design and preliminary evaluation. The channel emulator is based on an RFSoC device with four independent RF input/output (I/O) channels, which allows for the emulation of up to four unique wireless propagation channels simultaneously. Each I/O channel provides built-in analog-to-digital (ADC) and digital-to-analog (DAC) conversion to enable low-latency baseband processing of analog signals transmitted by connected RF devices, such as the MPSoC devices described in Sec.~3.2, via coaxial cable. Supported by the high-speed signal processing capabilities of the RFSoC FPGA fabric, this facilitates ``plug-and-play'' operation of C2Stack hardware devices without modification of network protocols or device reconfiguration. Figure 20 shows a snapshot of the HITL emulator.

Each emulated channel is modeled using a tapped delay line (TDL) filter with four taps to apply multi-path signal propagation characteristics, including path loss, phase offset, and Doppler/frequency shift for each path, defined by the delay and weight of each tap. Tap delays are fixed to match the symbol rate of the C2Stack hardware configuration discussed in Sec.\ref{sec:c2stack_module}. 
Tap weights are determined using an external ray tracing-based network simulation, which leverages NVIDIA Sionna \cite{hoydis2023sionna} to model transmission, reflection, refraction, and diffraction characteristics of a large number of propagation paths through a virtual model of the deployment environment and aggregate the resulting multi-path signal into a single complex-valued weight for each channel tap. After TDL filtering, signals are mixed according to the channel configuration to apply interference among signals designated to shared channels.

\end{document}